\documentclass{emulateapj-rtx4}
\usepackage{pdflscape,longtable}

\newcommand{\sdss}{{\small {SDSS}}}

\newcommand{\sdssii}{{\small {SDSS-II}}}
\newcommand{\sn}{{\small {SN}}}
\newcommand{\sne}{{\small {SN}}e}
\newcommand{\snia}{{\small {SN~I}}a}
\newcommand{\sneia}{{\small {SN}}e{\small{~I}}a}
\newcommand{\ugriz}{{\it {ugriz}}}
\newcommand{\galex}{{\small {\it{GALEX}}}}
\newcommand{\fuv}{{\it {FUV}}}
\newcommand{\nuv}{{\it {NUV}}}
\newcommand{\ukidss}{{\small {UKIDSS}}}
\newcommand{\yjhk}{{\it {YJHK}}}
\newcommand{\fsps}{{\small {FSPS}}}
\newcommand{\pegase}{{\small {P\'{E}GASE.2}}}

\newcommand{\salt}{{\small {SALT2}}}
\newcommand{\snana}{{\small {\tt SNANA}}}

\newcommand{\msun}{\mathrm{M}_\sun}
\newcommand{\musn}{{$\mu_{\mathrm{SN}}$}}
\newcommand{\muz}{{$\mu_z$}}
\newcommand{\pfit}{{$\mathcal{P}_\mathrm{fit}$}}
\newcommand{\smu}{{\small {SALT2mu}}}
\newcommand{\smupaper}{Marriner et al. 2011, accepted by ApJ}
\newcommand{\matpaper}{Smith et al. (2011), in prep}
\newcommand{\datarelpaper}{Sako et al., in prep}

\received{21 April 2011}
\accepted{29 July 2011}

\slugcomment{Accepted for publication in ApJ}

\shorttitle{Multi-wavelength Properties of \snia\ Host Galaxies}
\shortauthors{Gupta et al.}

\begin{document}

\title{Improved Constraints on Type Ia Supernova Host Galaxy Properties Using 
	Multi-wavelength Photometry and Their Correlations with Supernova Properties}

\author{
	Ravi~R.~Gupta\altaffilmark{1}, 
	Chris~B.~D'Andrea\altaffilmark{1}, 
	Masao~Sako\altaffilmark{1}, 
	Charlie~Conroy\altaffilmark{2} ,
	Mathew~Smith\altaffilmark{3}, 
	Bruce~Bassett\altaffilmark{4,5,6}, 
	Joshua~A.~Frieman\altaffilmark{7,8}, 
	Peter~M.~Garnavich\altaffilmark{9}, 
	Saurabh~W.~Jha\altaffilmark{10}, 
	Richard~Kessler\altaffilmark{7,11}, 
	Hubert~Lampeitl\altaffilmark{12}, 
	John~Marriner\altaffilmark{8}, 
	Robert~C.~Nichol\altaffilmark{12},
	and Donald~P.~Schneider\altaffilmark{13}
}

\email{ravgupta@physics.upenn.edu}

\altaffiltext{1}{\label{penn}
  Department of Physics and Astronomy,
  University of Pennsylvania, 209 South 33rd Street,
  Philadelphia, PA 19104, USA
}
\altaffiltext{2}{\label{cfa}
  Harvard-Smithsonian Center for Astrophysics,
  60 Garden Street, 
  Cambridge, MA 02138, USA
}
\altaffiltext{3}{\label{acgc}
  Astrophysics, Cosmology and Gravity Centre (ACGC),
  Department of Mathematics and Applied Mathematics,
  University of Cape Town, Rondebosch, 7701, South Africa
}
\altaffiltext{4}{\label{saao}
  South African Astronomical Observatory, 
  P.O. Box 9, Observatory 7935, South Africa
}
\altaffiltext{5}{\label{uct}
  Department of Mathematics and Applied Mathematics, 
  University of Cape Town, Rondebosch, 7701, South Africa
}
\altaffiltext{6}{\label{aims}
  African Institute for Mathematical Sciences, 
  Muizenberg, Cape Town, South Africa
}
\altaffiltext{7}{\label{uchi}
  Department of Astronomy \& Astrophysics, 
  University of Chicago, Chicago, IL 60637, USA
}
\altaffiltext{8}{\label{fnl}
  Fermilab, P.O. Box 500, Batavia, IL 60510, USA
}
\altaffiltext{9}{\label{nd}
  Department of Physics, 
  University of Notre Dame, Notre Dame, IN 46556, USA
}
\altaffiltext{10}{\label{rutgers}
  Department of Physics \& Astronomy, 
  Rutgers the State University of New Jersey, Piscataway, NJ 08854, USA
}
\altaffiltext{11}{\label{kicp}
  Kavli Institute for Cosmological Physics, 
  The University of Chicago, 5640 South Ellis Ave., Chicago, IL 60637, USA
}
\altaffiltext{12}{\label{icg}
  Institute of Cosmology and Gravitation, 
  University of Portsmouth, Portsmouth, PO1 3FX, UK
}
\altaffiltext{13}{\label{psu}
  Department of Astronomy \& Astrophysics, 
  The Pennsylvania State University, University Park, PA 16802, USA
}

\begin{abstract}
We improve estimates of stellar mass and mass-weighted average age of Type Ia supernova (\snia) host
galaxies by combining UV and near-IR photometry with optical photometry in our analysis.
Using 206 \sneia\ drawn from the full three-year \sdssii\ Supernova Survey 
(median redshift of $z \approx 0.2$) and multi-wavelength host-galaxy photometry from \sdss, \galex, 
and \ukidss, we present evidence of a correlation ($1.9 \sigma$ confidence level) between 
the residuals of \sneia\ about the best-fit Hubble relation and the mass-weighted average age of their 
host galaxies.  The trend is such that older galaxies host \sneia\ that are brighter than average after 
standard light-curve corrections are made.  We also confirm, at the $3.0 \sigma$ level, the trend seen by 
previous studies that more massive galaxies often host brighter \sneia\ after light-curve correction.  
\end{abstract}


\section{INTRODUCTION}

Observations of Type Ia supernovae (\sneia) are a key measurement in determining the standard 
cosmological model.  Their empirical luminosity-distance calibration based on relations between \snia\ peak 
luminosity and both light-curve width and optical colors \citep{phi93,ham96b,rie96} provides evidence for 
the accelerated expansion of the Universe and the existence of dark energy \citep{rie98,per99}.
According to the current theory, the progenitor of a \snia\ is a carbon-oxygen white dwarf that approaches 
the Chandrasekhar limit, resulting in a thermonuclear explosion \citep{whe73,hil00}.
However, the exact mechanism by which the progenitor accumulates this mass remains uncertain. 
Investigations of the physical properties of \snia\ host galaxies can provide insight into the 
environment in which these progenitor systems form.
Furthermore, although \sneia\ are remarkably standarizable, correcting for light-curve width and color still 
results in a scatter in peak brightness of $\sim 0.15$ mag \citep{guy07,jha07,con08}.  
Studying how variations in \snia\ luminosities depend on the environment of the progenitor will help 
reveal the origin of this scatter.

Over the years, several correlations between \sneia\ and the properties of their progenitors and environments 
have been discovered.  For example, intrinsically brighter \sneia\ tend to occur in galaxies with younger 
stellar populations while fainter ones often occur in passively evolving galaxies \citep{ham95,ham00,sul06}.
Studies have also shown that per unit stellar mass, the rate of occurrence of \sneia\ within a galaxy 
declines with decreasing SFR \citep{van90,man05,sul06}.  
In addition, properties of the progenitors themselves can directly influence light-curve properties of \sneia.  
Theoretical models generally agree that the metallicity of the white dwarf progenitor affects the amount of 
radioactive $^{56}$Ni produced in the thermonuclear explosion, the decay of which powers the light curve 
of the \sn\ \citep{hof98,tim03}.  Assuming that global metallicity correlates with progenitor metallicity, 
\citet{gal05} presented qualitative evidence suggesting that it is more likely that progenitor age, rather than 
metallicity, is primarily responsible for the variability in \snia\ peak luminosity.  
The true source of this variability has yet to be determined definitively.

More recently, \citet{gal08} found that early-type host galaxy metallicity is correlated with residuals on the \sn\ 
Hubble diagram around the best-fit cosmology.  The galaxy mass-metallicity relationship \citep{tre04} has led 
several authors to investigate whether mass is a proxy for this metallicity trend with Hubble residual (HR).  
Indeed, the latest studies have shown that more massive galaxies tend to host \sneia\ with residuals that are 
brighter than average after light-curve correction \citep{kel10,lam10,sul10}.  
Age is another host property that can be estimated, which might more directly influence \sn\ progenitor systems.
\cite{gal08} plotted HR against luminosity-weighted age using optical spectra from 29 early-type host galaxies 
but found no significant trend.  \citet{nei09} used optical and UV photometry to calculate luminosity-weighted 
ages of 166 nearby host galaxies.  They found that for the subsample of 22 low-extinction host galaxies, 
there was a $2.1 \sigma$ trend indicating that \sneia\ in older hosts have residuals that are brighter than 
average.  However, when the full sample was used, the trend disappeared.  

In this work, we use \snia\ host galaxy photometry spanning the ultraviolet, optical, and near-infrared bands, 
which allows us to constrain stellar masses and ages of host galaxies by comparing the observed photometry 
to synthetic photometry generated from stellar population synthesis models.  Knowledge of these physical 
properties of host galaxies can improve our understanding of \snia\ progenitors and the diversity of their 
light curves.

\section{DATA}

\subsection{Supernova Sample and Light Curve Analysis}
\label{LCcuts}

Our supernova sample consists of the spectroscopically confirmed \sneia\ discovered in the full three-year 
sample of the Sloan Digital Sky Survey (\sdssii) Supernova Survey \citep{fri08}.  These \sne\ lie in the 
redshift range 0.01 $<z<$ 0.42 with a median redshift of $z \approx 0.2$ and are located in Stripe 82, 
a 300 deg$^2$ equatorial strip of sky scanned repeatedly by \sdssii\ for three months a year from 2005 
to 2007 using a CCD camera on the \sdss\ 2.5 m telescope \citep{yor00,gun98,gun06}.  Over the course of
the Supernova Survey, $\sim 500$ \sne\ were spectroscopically confirmed to be Type Ia \citep{sak08,hol08}.
Unlike previous studies, we make use of the \sdss\ \sn\ sample over the entire redshift range for this work.

We use the publicly available Supernova Analysis package \snana\ \citep{kes09b} along with the \snana\ 
implementation of the light-curve fitter \salt\ \citep{guy07} to determine \sn\ properties for our sample 
based on the \sdssii\ photometry (\citealp{hol08}; \datarelpaper). We apply the following selection cuts to our sample, 
similar to those made in the cosmology analysis by \citet{kes09a}:
\begin{enumerate}  
	\item{At least one measurement with $T_\mathrm{rest}<-2$ days, where $T_\mathrm{rest}$ is the 
	rest-frame time, such that $T_\mathrm{rest}=0$ corresponds to peak brightness in rest-frame B band}
	\item{At least one measurement with $T_\mathrm{rest}>+10$ days}
	\item{At least one measurement with signal-to-noise ratio (S/N) $>$ 5 for each of the {\it g}, {\it r}, 
	and {\it i} bands}
	\item{\pfit\ $ > 0.001$, where \pfit\ is the \sneia\ light-curve fit probability based on 
	the $\chi^2$ per degree of freedom}
\end{enumerate}
These cuts reduce our sample size to 319 \sne.

\subsection{\sdss\ Host Galaxy Identification}

The \sdss\ contains photometric measurements in five optical passbands, \ugriz\ \citep{fuk96}. 
In order to match our \sne\ with host galaxies, we search the \sdss\ deep optical stacked images of Stripe 82 
\citep{dr7} for galaxies within a 0.25 arcminute radius of the \sn\ position, as was done by \citet{lam10} and 
\matpaper.  We choose the closest galaxy to be the host and require that the host \sdss\ model magnitude 
falls in the range $15.5 < r < 23$ to ensure robust photometry.
Of the 319 \sne\ that pass light-curve quality cuts, 14 (4\%) do not have identifiable hosts because they fall 
outside of the \sdss\ footprint, were too faint to be detected in the co-added images, or had $r$-band 
magnitudes outside our allowed range.  For the remaining 305 host galaxies, we visually confirm each match 
is correct by inspecting images with and without the \sn. In almost all cases, the host identification is 
unambiguous.  However, a spectroscopic redshift for both the \sn\ and the host galaxy is the only sure way 
guarantee a correct match, and this is the case for 80\% of the \sneia\ in the Supernova Survey.

\subsection{Host Matching and Galaxy Photometry}

Since our 305 host galaxies are \sdss-selected, we have \ugriz\ photometry for all hosts.  Nearly all 
magnitudes come from the Stripe 82 co-add catalogue, although for a few cases where the host is nearby and 
extended, deblending by the pipeline on the co-added image required that we use the DR7 \citep{dr7} 
catalogue magnitudes derived from single frames.  
We use the \sdss\ model magnitudes which are best for galaxy colors.
In addition to optical photometry, we obtain host photometry in the ultraviolet and 
near-infrared from the {\it Galaxy Evolution Explorer} (\galex) GR6 and the UKIRT Infrared Deep Sky 
Survey (\ukidss) DR5, respectively.  The \galex\ telescope images in two passbands, far-UV (\fuv) and 
near-UV (\nuv) \citep{mar05}. The \ukidss\ passbands are \yjhk\ and the photometric system is described 
in \citet{hew06}.  The description of the \ukidss\ survey is given in \citet{law07}.  
Model magnitudes, as defined by \sdss, are not computed by \galex\ and \ukidss.  Therefore,
we use Petrosian magnitudes \citep{pet76} for \ukidss\ and Kron-like elliptical aperture magnitudes 
\citep{kro80} for \galex\ since Petrosian magnitudes are not available in the \galex\ catalogue. 
The majority of galaxies in our sample are not large in angular size, and so the difference between these 
magnitudes should not be significant.
We exclude \ukidss\ objects which have been deblended because of a known error in the pipeline that 
results in erroneous Petrosian magnitudes for these objects \citep{smi09}.

Photometric data were obtained from online catalogues via SQL (Structured Query Language) queries 
through the \sdss\ Catalogue Archive Server (CAS)\footnote{\url{http://casjobs.sdss.org/CasJobs/}}, 
the \galex\ Multimission Archive at STScI (MAST) CAS\footnote{\url
{http://galex.stsci.edu/casjobs/}}, and the \ukidss\ WFCAM Science Archive 
(WSA)\footnote{\url{http://surveys.roe.ac.uk/wsa/}}.
The UV and near-IR data were obtained by cross-matching the \sdss\ host galaxy coordinates with 
the \galex\ and \ukidss\ catalogues using a 5\arcsec\ search radius.
Of the 305 \sdss\ host galaxies, 198 (65\%) have \galex\ matches and 178 (58\%) have \ukidss\ 
matches within 5\arcsec, while 127 (42\%) have matches in both \galex\ and \ukidss.

We do not require every galaxy to have photometry in all 11 bands (\fuv, \nuv, \ugriz\yjhk).  
The addition of UV data helps to constrain age, metallicity, and recent star formation, while near-IR 
data probe the older stellar populations that compose a large portion of the mass.  
For example, adding \galex\ data to \sdss\ data has been shown to greatly improve estimates of 
dust optical depth and star formation rate \citep{sal05}.

\section{METHODS}

\subsection{\sn\ Distance Modulus and Hubble Residuals}
\label{DMandHR}

The distance modulus for a particular \snia\ in the \salt\ model is given by 
\begin{equation}
\mu_{\mathrm{SN}} = m_B - M + \alpha x_1 - \beta c,
\label{eqnMu}
\end{equation}
where $x_1$ (stretch parameter), $c$ (color), and $m_B$ (apparent $B$-band magnitude at peak) are 
obtained from \salt\ for each \sn\ by fitting its light curve; $\alpha$ and $\beta$ are coefficients which we 
assume to be constant; and $M$ is the absolute magnitude.
The distance modulus along with $\alpha$ and $\beta$ are determined from the output of \salt\ using the 
program \smu\ (\smupaper), which is part of the \snana\ package.  \smu\ is able to calculate $\alpha$ and 
$\beta$ independent of cosmology by minimizing the scatter in the Hubble relation in small redshift bins.
Values of $\alpha$ and $\beta$ in this work are computed from the sample of \sdss\ \sneia\ that pass the 
light-curve cuts in Section \ref{LCcuts} and which are either spectroscopically-confirmed or 
photometrically-typed and have host redshifts.  We find the best-fit values to be $\alpha = 0.121$ and 
$\beta = 2.82$, and use these to obtain the distance modulus, \musn.
The Hubble Constant (which is degenerate with $M$) is effectively a constant offset to \musn\ and is an 
input to \smu; we choose $H_0 = 70$ km s$^{-1}$ Mpc$^{-1}$.

We define Hubble residuals as HR $\equiv$ \musn$-$\muz, where \musn\ is the distance modulus obtained 
from \sn\ light curves via \smu\ and \muz\ is the distance modulus calculated from the redshift of the \sn\ 
and the best-fit cosmology.  The best-fit cosmology here is determined by \salt\ based on the first-year 
\sdssii\ \sn\ sample \citep{kes09a}, i.e. $\Omega_\mathrm{M} = 0.274,\ \Omega_\Lambda = 0.735$.

A \sn\ with a HR $> 0$ signifies that it is fainter than expected for the best-fit cosmology even 
after correcting for light-curve shape.  Here it is useful to define ``underluminous" to refer to 
\sneia\ with HR $> 0$ and ``overluminous'' to refer to \sneia\ with HR $< 0$.
Errors in HR are derived by adding the errors on \musn\ and \muz\ in quadrature, where the errors on 
\muz\ are calculated as $[\mu(z+z_{err})-\mu(z-z_{err})]/2$.

\subsection{Galaxy Model Fitting}
\label{GalModFit}

Stellar population synthesis (SPS) codes are commonly used to create model templates of galaxies based on 
stellar evolution calculations with the goal of inferring galaxy properties such as mass, age, metallicity, and star 
formation.  We use the Flexible Stellar Population Synthesis code (\fsps\ v2.1) developed by \citet{con09} 
and updated in \citet{con10} to generate spectral energy distributions (SEDs) of composite stellar 
populations (CSPs). \fsps\ is similar to codes such as \citet{bc03} and \pegase\ \citep{fio97,leb02}, but has
increased flexibility in the initial mass function (IMF), dust model, and stellar evolution assumptions compared 
to other models \citep{con09}. 
For this work we use the BaSeL3.1 spectral library and the Padova isochrones as were used by \citet{con09}.
Since we are interested only in relative masses of our host galaxies and are not comparing masses directly 
with other works, the choice of IMF is not so important; here we adopt the commonly used \citet{cha03} IMF.
For details on \fsps\ and a comparison of spectral libraries, isochrones, and SPS codes, see 
\citet{con09} and \citet{con10}. 

Our models are generated on a grid of 4 \fsps\ parameters: 
metallicity, log$[Z/Z_\sun]$, assumed constant over time for each model; $\tau_{dust}$, dust 
attenuating old stellar light; $\tau_{SF}$, the $e$-folding time scale of star formation; and 
$t_{start}$, the time when star formation begins.  
The CSPs we use here each have exponentially declining star formation rates (SFRs), often called ``tau 
models" [SFR$(t) \propto \mathrm{exp}(-t/\tau_{SF})$], that we allow to be shifted in time by an amount 
$t_{start}$.  For each CSP, star formation is initiated at a time $t_{start}$ after the Big Bang and the 
rate of star formation declines exponentially thereafter, as dictated by $\tau_{SF}$.
We adopt the two-component dust model of \citet{cha00} in which the dust attenuation factor is 
exp$(-\tau_\lambda(t))$ and $\tau_\lambda(t)$ is the optical depth given by
\begin{equation}
\tau_\lambda(t) = \left\{
\begin{array}{ll}
\tau_{10}(\lambda/5500 \mathrm{\AA})^{-0.7} & t \leq \mathrm{10\ Myr} \\
\tau_{dust}(\lambda/5500 \mathrm{\AA})^{-0.7} & t > \mathrm{10\ Myr.} 
\end{array} \right.
\end{equation}
We fix $\tau_{10} = 3\tau_{dust}$, where $\tau_{10}$ is the optical depth of dust surrounding stars 
younger than 10 Myr and $\tau_{dust}$ is the optical depth of dust surrounding stars of greater age 
\citep{cha00,kon04,con09}. 
Table \ref{ParTable} lists the values of the \fsps\ parameters used to generate our model grid.  The limits on 
the grid values were chosen in an attempt to encompass reasonable values appropriate for the stellar populations 
of our host galaxy sample.  Our redshifts range from nearby to intermediate, indicating that our hosts 
are likely not extremely metal-poor.  The range on $\tau_{dust}$ is centered on the standard value given in 
\citet{cha00}.  A SFR with a $\tau_{SF}$ value of 0.1 Gyr closely resembles a single burst of star formation 
while a value of 10 Gyr is essentially a flat, constant SFR.  The maximum value of $t_{start}$ was chosen 
to be 7 Gyr after the Big Bang since it is unlikely that \emph{all} stars in a galaxy would be formed later than this.

\begin{deluxetable}{cc}
\tablecaption{\fsps\ model grid parameters \label{ParTable}}
\tablewidth{0pt}
\tablehead{
\colhead{\fsps\ parameter} & \colhead{Grid values}
}
\startdata
log$[Z/Z_\sun]$ & $-0.88, -0.59, -0.39, -0.20, 0, 0.20$\\
$\tau_{dust}$ &  $0, 0.1, 0.3, 0.5, 1.0, 1.5$\\
$\tau_{SF}$ (Gyr) &  $0.1, 0.5, 1, 2, 3, 4, 6, 8, 10$\\
$t_{start}$ (Gyr) & $0, 1, 2, 3, 4, 5, 6, 7$\\
\enddata
\end{deluxetable}

The models produce photometry in \fuv, \nuv, \ugriz, and \yjhk\ for direct comparison to observed data 
from \galex, \sdss, and \ukidss.  The spectroscopic redshift of the \sn\ is used to obtain the synthetic 
apparent magnitudes for each model SED. In calculating derived galaxy properties, we assume the 
aforementioned \citet{kes09a} cosmology ($\Omega_\mathrm{M} = 0.274, \Omega_\Lambda = 0.735$) 
along with $H_0 = 70$ km s$^{-1}$ Mpc$^{-1}$, for consistency.  
Our results are not strongly affected by our choice of cosmology.

All magnitudes are corrected for Milky Way extinction using the maps of dust IR emission from 
\citet{sfd98} in conjunction with the extinction curve of \citet{car89}. 
The \sdss\ and \ukidss\ magnitudes are then corrected to the AB system \citep{oke83}, using 
\citet{kes09a} and \citet{hew06}, respectively.  
We add minimum calibration errors from \citet{bla03} in quadrature to all \sdss\ magnitude errors
(0.05, 0.02, 0.02, 0.02, and 0.03 mag for \ugriz, respectively) to account for systematic effects.  
For \galex\ and \ukidss\ we add a minimum calibration error of 0.02 mag in quadrature with the 
photometric error for each band as well.
All magnitudes and errors are converted to flux.  A least-squares fit is then performed in flux between 
the data and each of the model SED fluxes, taking into account the photometric errors.  

In analogy to the $\chi^2$ cuts performed on the \sne\ sample, we remove any galaxies for which
the probability of the data being drawn from the best-fit model is $< 0.001$.  This criterion removes 
one third of our hosts from our sample and brings the final \sn-host sample size to 206.
This is the sample we will examine for this study.

\subsection{Derived Galaxy Properties}
\label{DerGalProps}

From the fit parameters for each SED model we derive two physical properties of our host galaxies: 
stellar mass and mass-weighted average age. 
Stellar mass (mass currently in stars) is calculated by multiplying the observed, de-reddened luminosity in 
the {\it r} band by the model mass-to-light ratio in the same band.  
The mass-weighted average age of the galaxy is computed as
\begin{equation}
\langle \mathrm{Age} \rangle = A - \frac{\int_0^A{t\Psi(t)dt}}{\int_0^A{\Psi(t)dt}},
\end{equation} 
where $A$ is the age of the Universe at the redshift of the \sn\ minus $t_{start}$ and 
$\Psi(t)$ is the SFR as a function of time in units of $\msun\ \mathrm{yr}^{-1}$.
For each galaxy, we calculate the median mass and age and the corresponding 68\% confidence intervals 
around the median (analogous to a $\pm1 \sigma$ range for a Gaussian distribution).  These uncertainties 
are obtained from the probability density functions (PDFs) constructed for both mass and age from the 
likelihoods of the models where each model is given a weight $\propto \mathrm{exp}(-\chi^2/2)$.
We take this PDF to be a sampled version of the true continuous distribution (which may not be Gaussian).  
In this way, our mass and age estimates are marginalized over the \fsps\ parameters such as metallicity and dust.
In Table \ref{DataTable} we list the \sne\ used in our final sample, the host galaxy coordinates, 
the redshift of the \sn, the host galaxy stellar mass and mass-weighted age, the \salt\ color and stretch 
parameters, and the HR.
A complete list of the \sne\ from years two and three of the \sdssii\ Supernova Survey along with photometry 
and other associated data will be published in \datarelpaper.

\section{RESULTS}

\subsection{Host Galaxy Properties}

\begin{figure}[tbp]
\epsscale{1.2}
\plotone{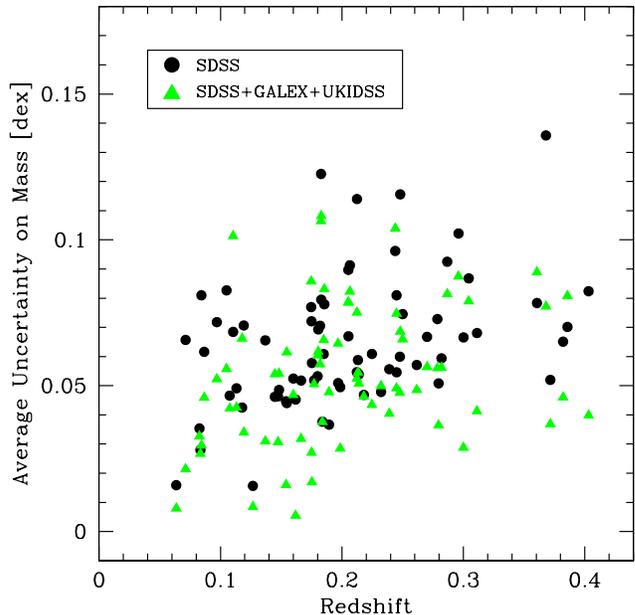}
\caption{Average mass uncertainty as a function of redshift for the sample of 71 galaxies which have 
photometry in optical, UV, and near-IR.  Black circles indicate results obtained from fits using \sdss\ data 
only; triangles indicate results obtained from fits using \sdss, \galex, and \ukidss\ data.}
\label{figMassErrvz}
\end{figure}

\begin{figure*}[tbp]
\epsscale{1.15}
\plottwo{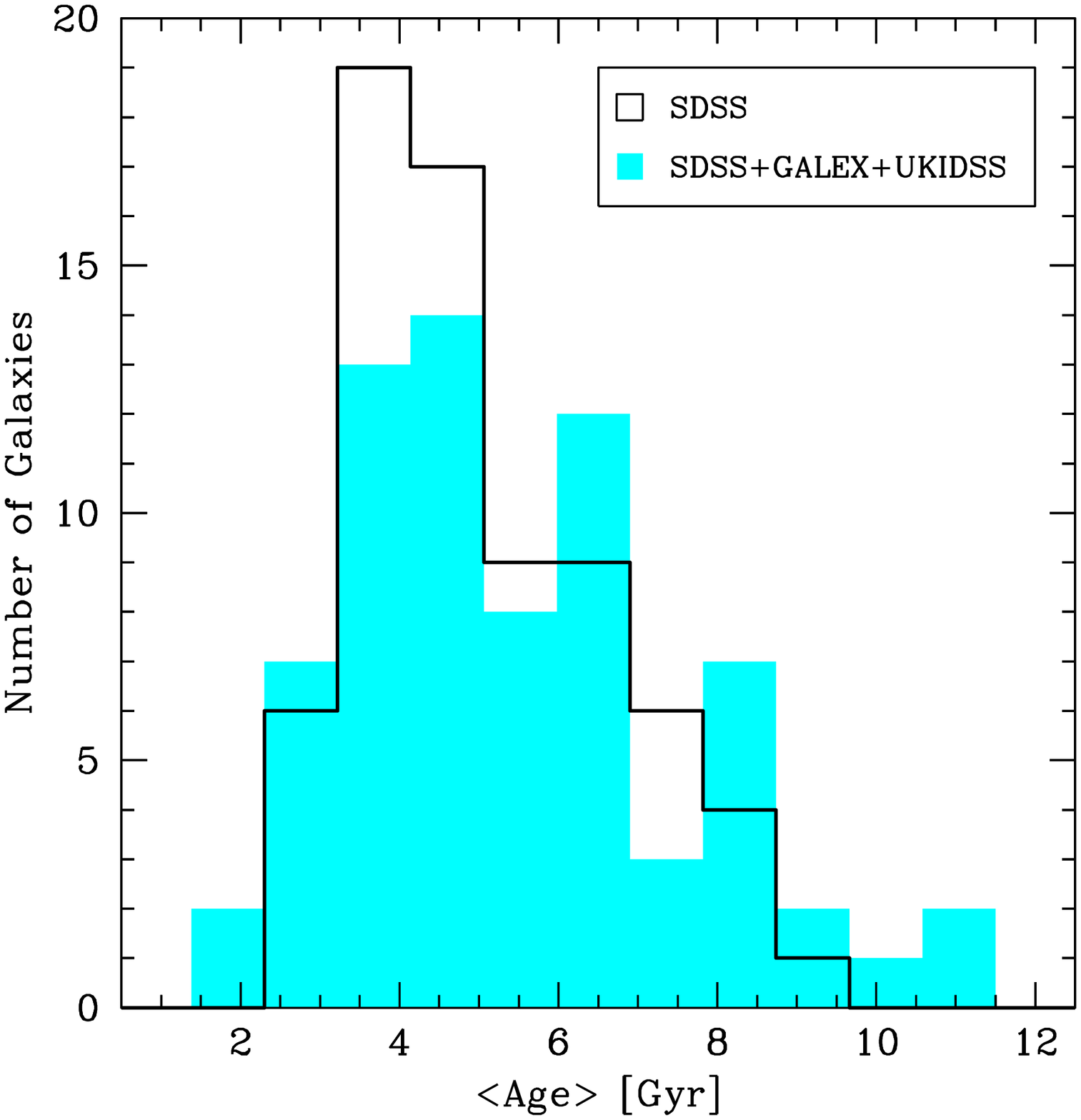}{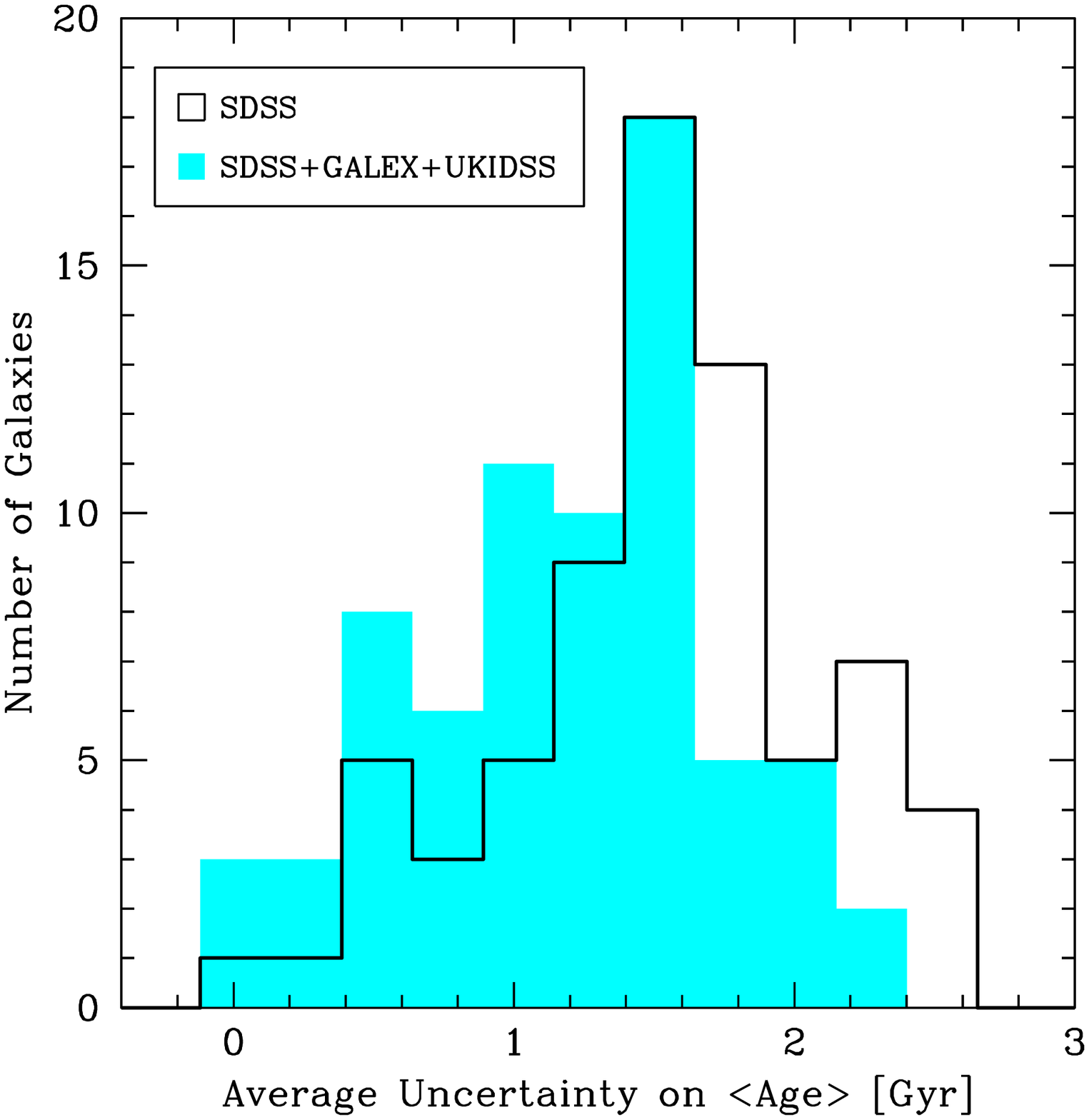}
\caption{\emph{Left}: Distributions of mass-weighted average age for the sample of 71 galaxies which have 
photometry in optical, UV, and near-IR showing the effect of adding \galex\ and \ukidss\ data to \sdss\ data.  
\emph{Right}:  Distributions of the average uncertainty on the age for the same sample.}
\label{figHistAge}
\end{figure*}

To determine if adding UV and near-IR photometry to optical data improves constraints on physical properties 
of our host galaxies, we examine the sample of 71 \sdss\ galaxies that have matches in both \galex\ and 
\ukidss\ (after all cuts are made).  
We find that while adding \galex\ and \ukidss\ data to the \sdss\ observations does not significantly 
change our resulting host masses, it does reduce the average uncertainties in the mass estimates (see 
Figure~\ref{figMassErrvz}), where the average uncertainty here is the mean of the upper and lower $1 \sigma$ 
uncertainties.  The uncertainty in mass increases with redshift because the photometric errors increase with 
redshift, but adding UV and near-IR data reduces these uncertainties in mass overall by 17\%.  
The addition of UV and near-IR data widens the range of the host age distribution while also 
reducing the average uncertainty in age on the whole by 22\% (see Figure \ref{figHistAge}).

\begin{figure}[tbp]
\epsscale{1.2}
\plotone{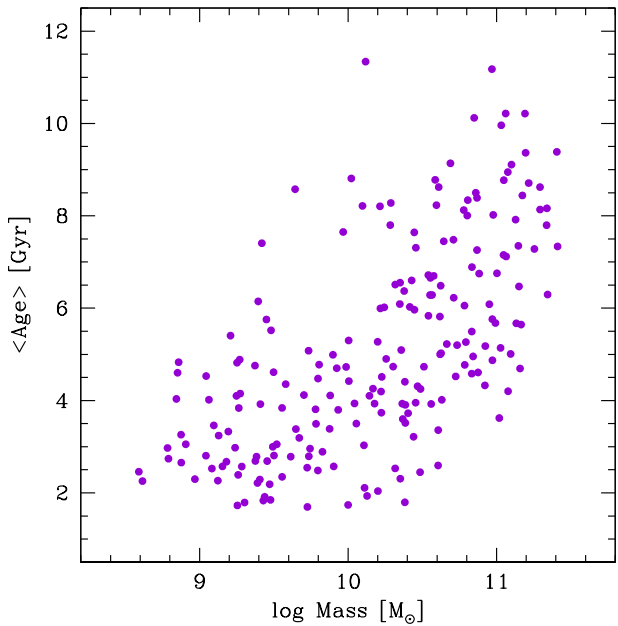}
\caption{Mass-weighted average age as a function of the stellar mass for our  
host galaxies.  As one would expect, the more massive galaxies tend to be the older galaxies, 
though the scatter in the relation is large.}
\label{figAgevMass}
\end{figure}

Figure \ref{figAgevMass} shows a plot of mass-weighted average age versus the stellar mass of 
our sample of host galaxies.  The distribution exhibits the expected trend that, in general, the most massive 
galaxies are also the oldest.  However, there appears to be an absence of low-mass old galaxies.  
This may be due to several factors, one of which is that for a given mass, older 
galaxies will be harder to detect by \sdss\ because they are fainter in the optical due to a dearth of young,
bright stars.  This absence of small, old galaxies may also be due to the fact that these galaxies likely
have a low SFR per unit mass and therefore do not produce many Type Ia events \citep{van90,man05,sul06}.

\subsection{Correlations with SN Fit Parameters}

\begin{figure*}[tbp]
\epsscale{1.15}
\plottwo{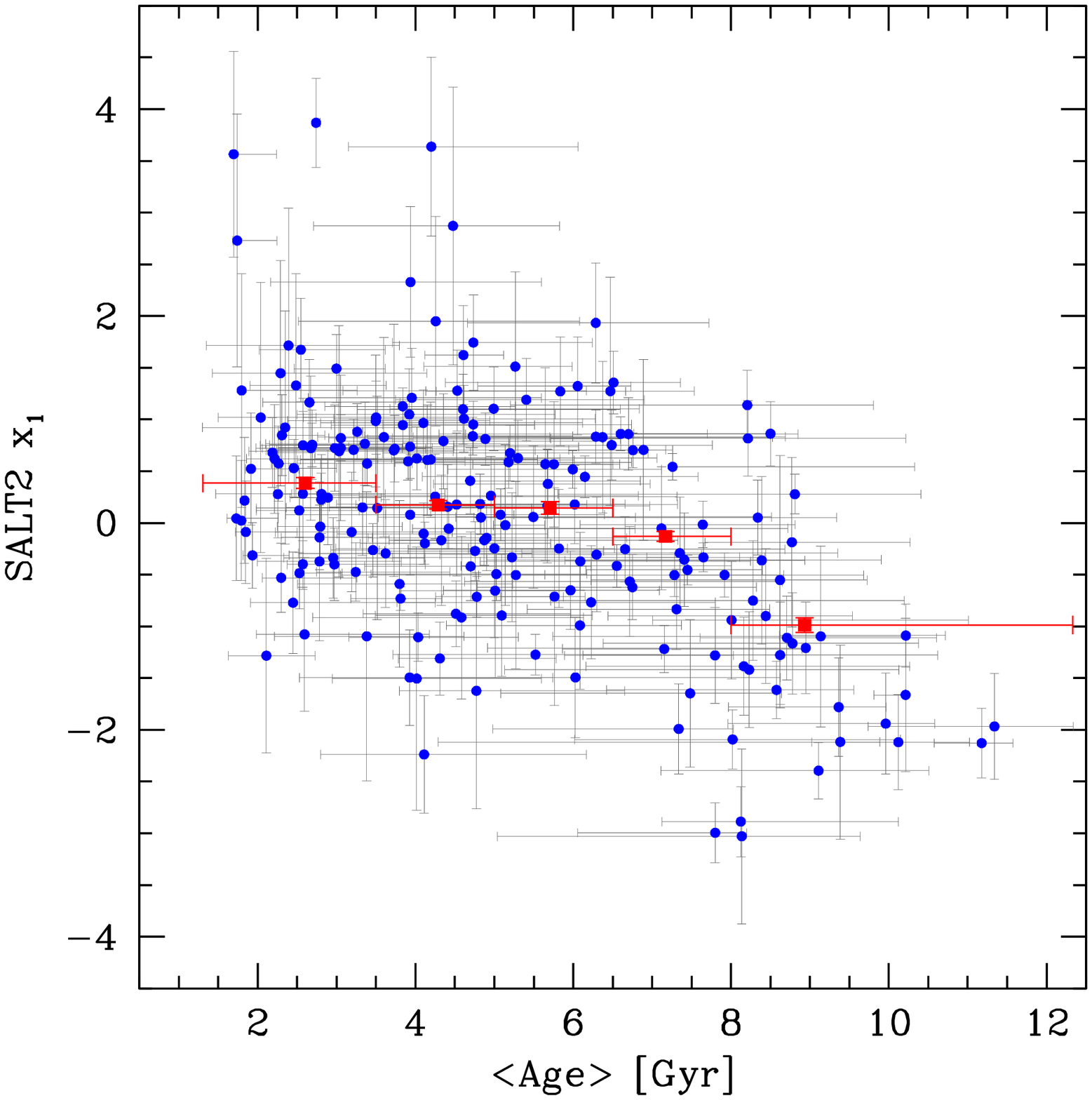}{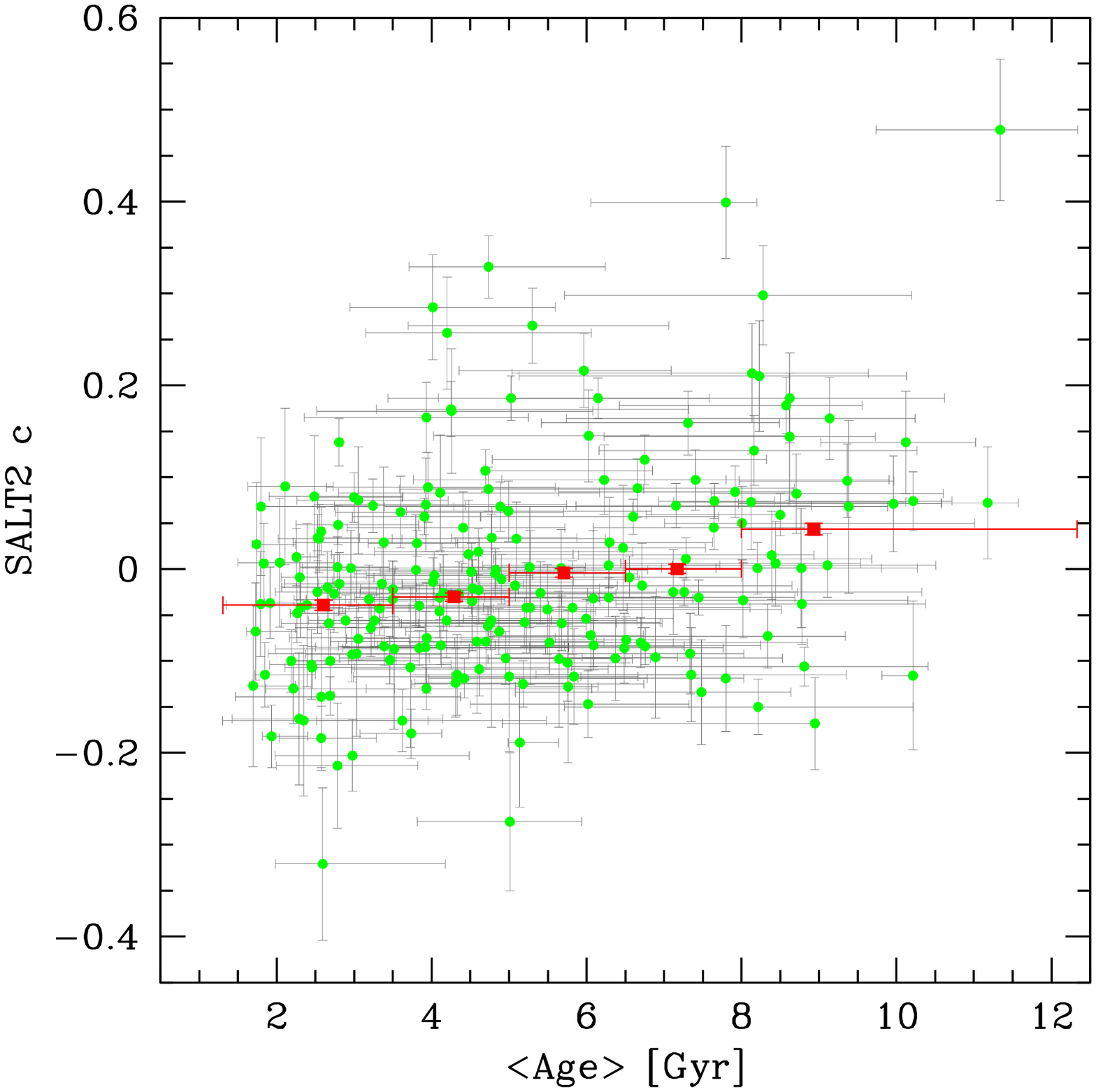}
\caption{\salt\ stretch ($x_1$) and color ($c$) as a function of mass-weighted average age 
of the host galaxy.  
The squares are binned averages calculated by taking the mean age and the inverse 
variance-weighted mean \salt\ parameter in each bin.  
The errors bars show the size of the bin (1.5 Gyr) and the 
$1\sigma$ error on the mean \salt\ parameter.}
\label{figSALTvAge}
\end{figure*}

Figure \ref{figSALTvAge} plots the \salt\ \sn\ fit parameters, stretch and color, as a function of host 
galaxy mass-weighted average age.  By definition, higher values of stretch correspond to intrinsically brighter 
\sneia.  Our results indicate that intrinsically brighter \sne\ occur preferentially in younger stellar populations.
This is consistent with the known trend that brighter \sne\ occur in late-type \citep{ham96a,gal05}, 
star-forming galaxies \citep{sul06}, and in bluer environments \citep{ham00}, since these types of 
galaxies are generally also young.
The trend we see of \sn\ color as a function of host age is not as clear; the distribution is essentially flat, 
although extreme values of color do seem to correlate with age.  However, since the \salt\ $c$ parameter 
encapsulates not only intrinsic \sn\ color but also possible extinction due to dust in the host galaxy, 
a definitive statement cannot be made about the relation between \sn\ color and host age.
Plots of stretch and color versus the host mass are not shown here, though our results strongly resemble 
those found in \citet{how09}, \citet{nei09}, and \citet{sul10}.

\subsection{Linear Trends with Hubble Residuals}
\label{LinTrends}

Linear regression has a long history in astronomy where there are often measurement errors in both the 
``dependent" and ``independent" variables.  There is, however, no concensus on the best method to use 
when fitting a line.  
Here, we fit for a linear dependence of HR with age and mass using the package LINMIX \citep{kel07}, as 
was used to determine the significance of trends with HR by \citet{kel10}.  LINMIX is a Bayesian approach to 
linear regression using a Markov chain Monte Carlo (MCMC) analysis, assuming that the measurement errors 
are Gaussian.  We make the assumption that our errors on the host properties are Gaussian and input into 
LINMIX the average of the upper and lower $1 \sigma$ uncertainties as the error in the dependent variable.  

When fitting, we do not add the intrinsic uncertainty (0.14 mag for \salt) in quadrature to the HR errors 
that is added by others when fitting for trends of host properties with HR \citep{kel10,sul10,lam10}.
This intrinsic uncertainty arises from the fit to the Hubble diagram and is the amount of scatter that must 
be added to the distance modulus such that the reduced $\chi^2$ of the best-fit cosmology is close to unity.
This is done in an attempt to account for unknown effects on \snia\ luminosity by factors not accounted for 
in the light-curve correction process, e.g. the properties of the host galaxy.  The effect of host galaxy 
properties on \snia\ is precisely the purpose of our study, so including the intrinsic scatter has the effect of 
weakening the strength of the measured correlations.
If we perform the fit including the intrinsic uncertainty, we find our best-fit slopes and intercepts 
vary only slightly, but the significances of the non-zero slopes drop by about $0.2 \sigma$.

\begin{figure}
\epsscale{1.2}
\plotone{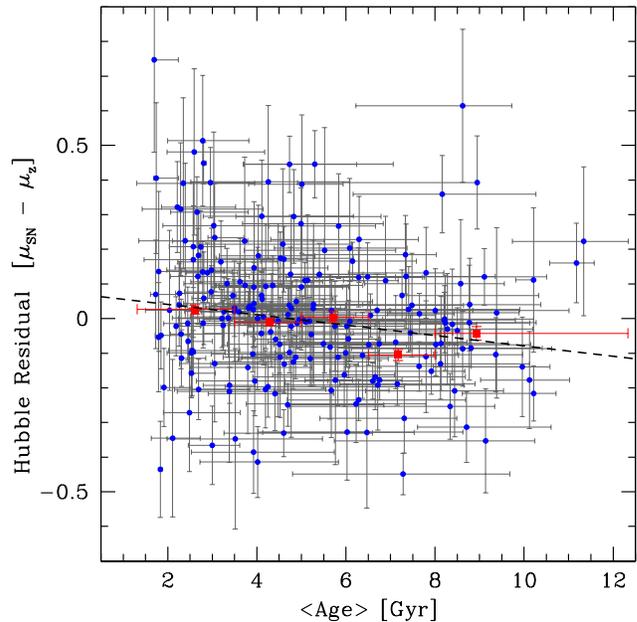}
\caption{Hubble residual as a function of mass-weighted average age of the host galaxy.
The squares are binned averages calculated by taking the mean age and the inverse 
variance-weighted mean HR in each bin.  The errors bars show the size of the bin (1.5 Gyr) 
and the $1\sigma$ error on the mean HR.  
The overplotted line shows the best fit to all the data points as described in Section \ref{LinTrends} 
and is given by the equation
HR = $-0.015 \times \langle \mathrm{Age} \rangle + 0.071 $. 
Of the MCMC realizations, 2\% have a slope greater than zero, and the significance of the deviation of
the best-fit slope from zero is $1.9 \sigma$.  
}
\label{figHRvAge}
\end{figure}

\begin{figure}
\epsscale{1.2}
\plotone{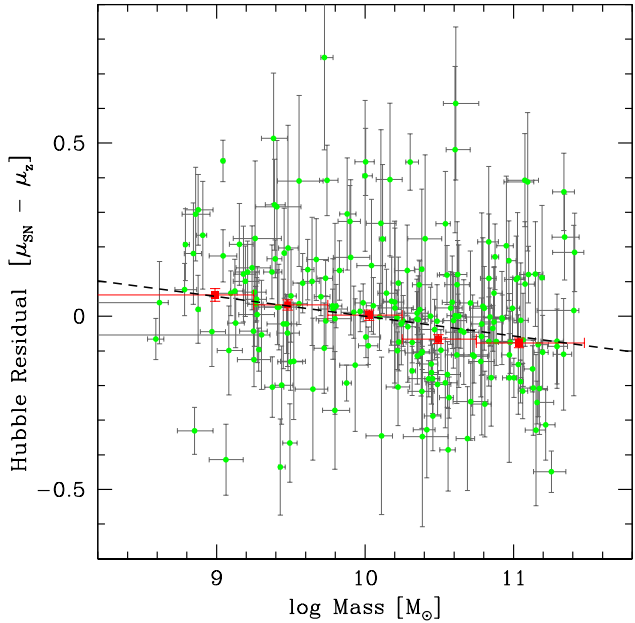}
\caption{Hubble residual as a function of stellar mass of the host galaxy.
The squares are the binned averages as described in Figure \ref{figHRvAge} and the bin size is 0.5 dex.
The overplotted line shows the best fit to all the data points as described in Section \ref{LinTrends} 
and is given by the equation
HR = $-0.057 \times \mathrm{log M} + 0.57 $.
Of the MCMC realizations, 0.1\% have a slope greater than zero, and the significance of the deviation of
the best-fit slope from zero is $3.0 \sigma$.}
\label{figHRvMass}
\end{figure}

In Figure \ref{figHRvAge}, we plot HR versus the mass-weighted average age of the host galaxy.
Figure~\ref{figHRvMass} shows HR versus the stellar mass of the host galaxy.
The overplotted lines are the best-fit model as determined from LINMIX.  
In all our LINMIX analyses we use 100,000 MCMC realizations.
For the HR trend with age we find the equation of the best-fit line to be 
\begin{equation}
\mathrm{HR} = -0.015(\pm 0.008) \times \langle \mathrm{Age} \rangle + 0.071(\pm 0.038).
\end{equation} 
The MCMC realizations in LINMIX are used to generate a sampling of the posterior distribution on the slope.
Of the MCMC realizations, 2\% have a slope greater than zero.  Fitting a Gaussian to the posterior slope 
distribution yields a mean of $-0.015$ and a standard deviation of 0.008.  
Based on this Gaussian fit, the mean slope differs from a slope of zero by $1.9 \sigma$.
Thus, for the HR-age correlation we quote the significance of a non-zero slope as $1.9 \sigma$.
For the HR trend with mass the best-fit line is
\begin{equation}
\mathrm{HR} = -0.057(\pm 0.019) \times \mathrm{log M} + 0.57(\pm 0.19) . 
\end{equation}
Of the MCMC realizations, 0.1\% have a slope greater than zero.  This corresponds to a $3.0 \sigma$ 
significance of a non-zero slope.

Our results indicate that after light-curve correction, there appears to be a deficit of underluminous \sne\ 
in older, more massive galaxies.  To test whether this result is due to incompleteness, we investigated the 
subsample of 40 \sneia\ for which \sdss\ is complete ($z \leq 0.15$).  Up to $z = 0.15$ the \sdssii\ \sn\ 
survey is estimated to be $\sim$ 100\% efficient for spectroscopic measurement \citep{kes09a}, so any 
\snia\ that may have occurred should have been detected in this subsample once the host is subtracted 
from the image.  Applying the method used on the full sample to the complete subsample reveals that the 
trend of HR with mass persists, with a slightly increased significance of $3.4 \sigma$.  
However, the HR trend with age is not statistically significant ($1.2 \sigma$) for the complete subsample.

\subsection{\sdss\ Co-add vs. Single-Frame Photometry}
\label{CoaddVsingle}

We also performed our analysis using \sdss\ Petrosian magnitudes from the DR7 catalogue, which are 
derived from the single frame images, in place of the Stripe 82 catalogue co-add model magnitudes, 
which are derived from the stacked images.  
A comparison of single-frame Petrosian magnitudes with co-add model magnitudes shows the two types 
of magnitudes agree for the most part, though there is scatter in the difference which increases with 
magnitude.  The $u$-band difference exhibits the largest scatter and a slight bias indicating that the 
single-frame Petrosian $u$-band magnitudes tend to be brighter.

We find that using single-frame photometry can change the derived galaxy properties and the 
uncertainties on these properties, thus possibly affecting the significance of trends with HR.  
The photometric errors on the single-frame magnitudes are roughly ten times larger than the photometric 
errors on the co-add magnitudes.  As a result, using single-frame magnitudes reduces the $\chi^2$ values 
of the SED fits and expands the range of SED models that provide reasonable fits, according to our method 
described in Section \ref{DerGalProps}.  This then has the effect of potentially shifting the median and 
increasing the width of the $\chi^2$-weighted PDF for the galaxy properties, which changes our derived 
values for these properties and increases their uncertainties.  
The derived galaxy mass is robust and relatively unaffected by the difference between single-frame and 
co-add photometry, although the uncertainties on the mass are more than twice as large for the single-frame 
data.  The average age is much more sensitive to this difference.  Overall, the single-frame Petrosian 
magnitudes produce younger ages and larger uncertainties on the age, but the scatter in both of these 
quantities is large.  The younger ages may be due in part to the $u$-band magnitude difference, since 
more flux in bluer bands can be interpreted as light from younger stars.
As a test, we inflated the errors on the co-add magnitudes by a factor of 10 and found 
that the results essentially reproduce those obtained from using the single-frame Petrosian magnitudes 
suggesting that the size of the photometric errors plays a substantial role in the discrepancy in derived 
galaxy properties.

\section{DISCUSSION}
\label{Discussion}

We confirm with a significance of $3.0 \sigma$ the result found by \citet{kel10}, \citet{sul10}, and 
\citet{lam10} that massive galaxies tend to host overluminous \sneia.  We also find indication, with a 
significance of $1.9 \sigma$, that even after light-curve correction, overluminous \sneia\ tend to occur 
in older stellar populations.  
We note that the \citet{nei09} trend of HR with host age was based on luminosity-weighted age while 
in this work we calculate mass-weighted age, making a direct comparison difficult.  However, the 
direction of the age trend we see agrees with the \citet{nei09} trend for their low-extinction hosts.  
The HR trend with host luminosity-weighted age plotted by \citet{gal08} is in the opposite direction from 
the trend we find here, though the significance of their trend is negligible and their methods different. 
We expect that mass-weighted age is a more unbiased measure of the age of the galaxy because it 
is not as strongly affected by UV flux from young stars as luminosity-weighted age \citep[see][Table 1]
{lee07}.  Furthermore, the mass-weighted age gives more weight to older stellar populations, 
to which \snia\ progenitors most probably belong, and is therefore more likely to be correlated with \snia\ 
properties than luminosity-weighted age.  

\begin{figure*}[tbp]
\epsscale{1.15}
\plottwo{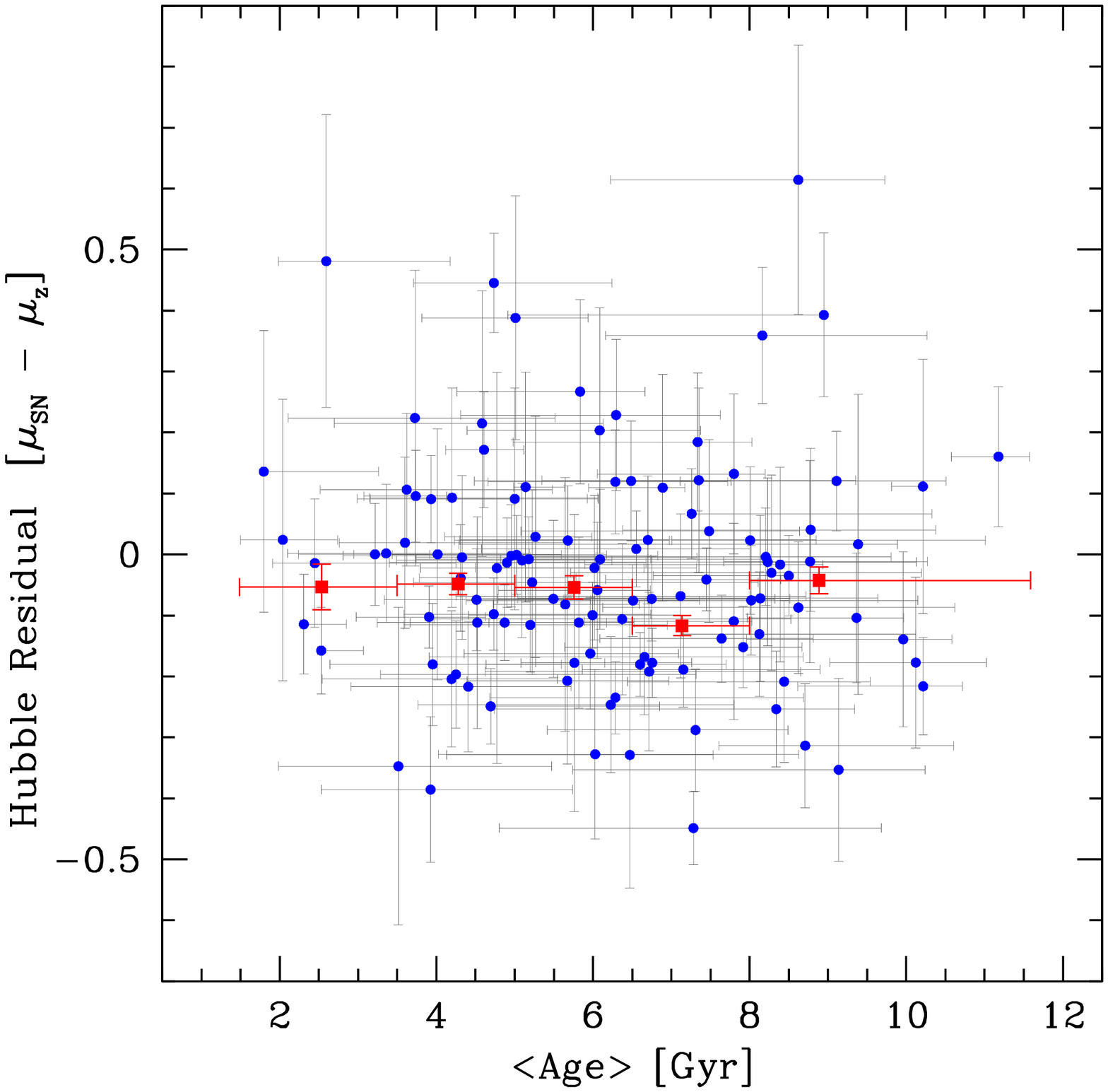}{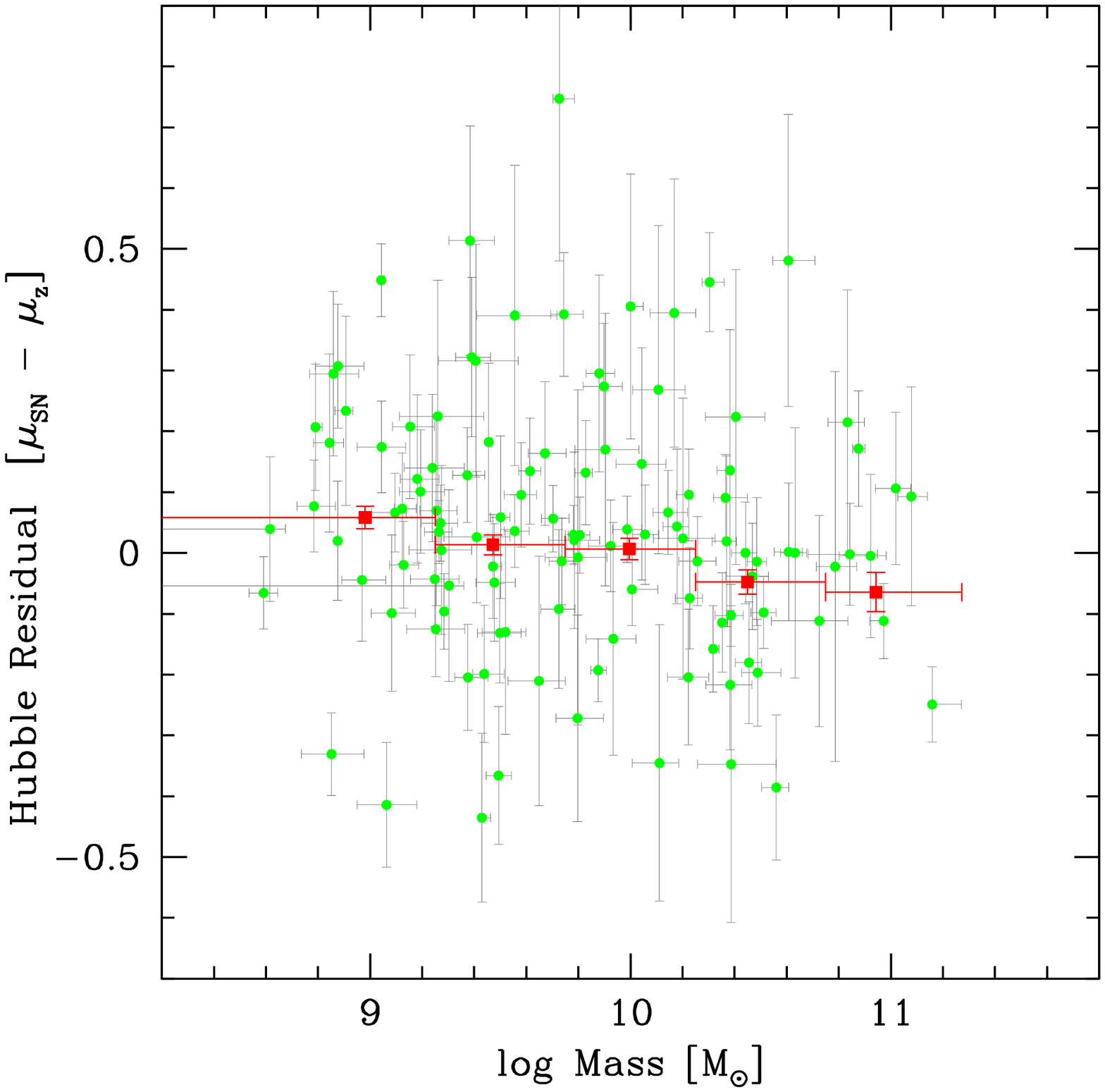}
\caption{\emph{Left}: HR versus host galaxy age for the logMass $>10.2$ group.
\emph{Right}:  HR versus host galaxy mass for the $\langle \mathrm{Age} \rangle <5$ Gyr group.
The squares are the binned averages.}
\label{figControl}
\end{figure*}

The trends we find of HR with mass and age agree with each other in the sense that galaxy mass and age 
are correlated, with older galaxies generally being more massive.  Based on the mass-age distribution in 
Figure~\ref{figAgevMass}, we split our data into two groups: an $\langle \mathrm{Age} \rangle<5$ Gyr 
group (which encompasses nearly the entire range of masses) and a logMass $>10.2$ group (which 
encompasses nearly the entire range of ages).  This was done in an effort to investigate the effect of one 
of the variables (mass or age) on HR while attempting to ``control" for the other.  In Figure~\ref{figControl} 
we plot HR against age for the logMass $>10.2$ group and HR against mass for the $\langle \mathrm{Age} 
\rangle <5$ Gyr group.   Within these groups we find that the HR correlation with mass is much weaker than 
in the full sample, and that the HR plot with age is consistent with no correlation.  
Estimates of host galaxy mass are more robust and have smaller uncertainties than estimates of 
age\footnote{Our method yields uncertainties of $2\%$ for mass and 27\% for age, though we emphasize 
that these uncertainties are \emph{statistical} only (as described in Section \ref{DerGalProps} and 
discussed in Section \ref{CoaddVsingle}) and that systematic uncertainties on mass are around 0.1 dex 
(25\%) at best.  Running our linear regression on the HR vs. mass plot with mass uncertainties inflated to 
be at least 0.1 dex does not change our results.}, and this in part may be why HRs are more strongly 
correlated with mass.  Thus, the true underlying property influencing the \snia\ explosion is still unclear.  
The strength of both correlations may be improved by having UV and near-IR matches for all \sdss\ 
hosts and by calculating galaxy magnitudes in a consistent manner through matched apertures for all 
survey types (as was done, e.g., in \citealp{hil10}), ensuring that UV-optical-NIR colors are accurate.  
It is also possible that the relationship between HR and age or mass is not simply linear and may be 
more complex.
Additionally, in an ideal case of a well-resolved extended host, a local age computed from photometry 
obtained from the location of the \sn\ in the galaxy would be preferable to the global galaxy average age 
that we compute here and would likely correlate more strongly with properties of the \snia.

We find that the HR trend with mass persists if we consider the subsample for which \sdss\ is complete.  
Therefore, based on our measurements and the completeness of our dataset, it is likely that \sneia\ 
that are underluminous after light-curve correction do not occur in massive galaxies.
This view is consistent with the results of \citet{kas09} who showed using \snia\ simulations and 
the \citet{tim03} model that metallicity can affect the explosion physics in such a way as to cause 
metal-rich progenitors to produce \sneia\ that are fast-declining and intrinsically fainter at peak.  
The usual light-curve correction technique does not account for this metallicity effect, and so metal-rich 
progenitors result in overluminous \sneia\ after corrections for light-curve shape.  Thus the \citet{kas09} 
result is in agreement with the trends we see with mass (and, for the full sample, age) since galaxies with 
higher metallicity are, in general, older and more massive \citep{tre04,glz05}.

Given our dataset and the large scatter in our trends of HR with age and mass, it is not possible to 
determine for certain what host property most influences \snia\ luminosity.  It is improbable that host 
mass itself, though better estimated, has a direct impact on \sneia.  Rather, it is more likely that host 
mass is correlated with other properties of the host that do directly influence the progenitors of \sneia.
The complexity of the relationships between galaxy properties such as age, mass, metallicity, dust, and 
SFR makes disentangling the factors that affect \sneia\ a challenge.
Further study is needed to truly ascertain the origin of these correlations between host properties 
and Hubble residuals and to potentially pinpoint the cause of these observed trends.

\acknowledgments

We are grateful to Jennifer Mosher and John Fischer for their advice and suggestions.  
We also thank C. Jonathan MacDonald for his programming insight, Bernie Shiao (STScI) 
for his help with \galex\ photometry, Steve Warren \& Daniel Mortlock (\ukidss) for their help in 
understanding the \ukidss\ photometry, and Patrick Kelly for his assistance with linear fitting.

\galex\ (Galaxy Evolution Explorer) is a NASA Small Explorer, launched in 2003 April.  
We gratefully acknowledge NASA's support for construction, operation, and science analysis for 
the \galex\ mission, developed in cooperation with the Centre National d'\'{E}tudes Spatiales of France 
and the Korean Ministry of Science and Technology.

This work is based in part on data obtained as part of the UKIRT Infrared Deep Sky Survey (\ukidss).
The United Kingdom Infrared Telescope (UKIRT) is operated by the Joint Astronomy Centre on behalf 
of the Science and Technology Facilities Council of the U.K.

Funding for the creation and distribution of the \sdss\ and \sdssii\ has been provided by the Alfred 
P. Sloan Foundation, the Participating Institutions, the National Science Foundation, the U.S. 
Department of Energy, the National Aeronautics and Space Administration, the Japanese Monbukagakusho, 
the Max Planck Society, and the Higher Education Funding Council for England. The SDSS Web site is 
\url{http://www.sdss.org/}.

The SDSS is managed by the Astrophysical Research Consortium for the Participating Institutions. 
The Participating Institutions are the American Museum of Natural History, Astrophysical Institute 
Potsdam, University of Basel, Cambridge University, Case Western Reserve University, University 
of Chicago, Drexel University, Fermilab, the Institute for Advanced Study, the Japan Participation 
Group, Johns Hopkins University, the Joint Institute for Nuclear Astrophysics, the Kavli Institute for 
Particle Astrophysics and Cosmology, the Korean Scientist Group, the Chinese Academy of Sciences 
(LAMOST), Los Alamos National Laboratory, the Max-Planck-Institute for Astronomy (MPA), the 
Max-Planck-Institute for Astrophysics (MPIA), New Mexico State University, Ohio State University, 
University of Pittsburgh, University of Portsmouth, Princeton University, the United States Naval 
Observatory, and the University of Washington.

This work is based in part on observations made at the following telescopes. The Hobby-Eberly Telescope 
(HET) is a joint project of the University of Texas at Austin, the Pennsylvania State University, Stanford 
University, Ludwig-Maximillians-Universit\"{a}t M\"{u}nchen, and Georg-August-Universit\"{a}t G\"{o}ttingen. 
The HET is named in honor of its principal benefactors, William P. Hobby and Robert E. Eberly. The Marcario 
Low-Resolution Spectrograph is named for Mike Marcario of High Lonesome Optics, who fabricated several 
optical elements for the instrument but died before its completion; it is a joint project of the Hobby-Eberly 
Telescope partnership and the Instituto de Astronom\'{i}a de la Universidad Nacional Aut\'{o}nomade 
M\'{e}xico.  The Apache Point Observatory 3.5 m telescope is owned and operated by the Astrophysical 
Research Consortium.  We thank the observatory director, Suzanne Hawley, and site manager, Bruce Gillespie, 
for their support of this project. The Subaru Telescope is operated by the National Astronomical Observatory 
of Japan.  The William Herschel Telescope is operated by the Isaac Newton Group on the island of La Palma in 
the Spanish Observatorio del Roque de los Muchachos of the Instituto de Astrofisica de Canarias.  The W. M. 
Keck Observatory is operated as a scientific partnership among the California Institute of Technology, the 
University of California, and the National Aeronautics and Space Administration; the observatory was made 
possible by the generous financial support of the W. M. Keck Foundation.

This work was funded by the NASA ADP Program NNX09AC75G.

{\it Facilities:} \facility{Sloan}, \facility{GALEX}, \facility{UKIRT}.

\clearpage 
\LongTables
\begin{landscape}
\begin{deluxetable}{llccccccccccccc}
\tabletypesize{\scriptsize}
\tablecaption{Properties of \sneia\ Sample and Host Galaxies \label{DataTable}}
\tablecolumns{14}
\tablewidth{0pt}
\tablehead{
\multicolumn{2}{c}{Designation} &
\multicolumn{2}{c}{Host Coordinates} &
\colhead{Redshift \tablenotemark{a}} &
\colhead{$M-$} &
\multicolumn{1}{c}{$M$} &
\colhead{$M+$} &
\colhead{$\mathrm{Age}-$} &
\multicolumn{1}{c}{Age} &
\colhead{$\mathrm{Age}+$} &
\colhead{$c$} &
\colhead{$x_1$} &
\colhead{HR}\\
\colhead{SN ID} &
\colhead{IAU} &
\colhead{$\alpha(J2000)$} &
\colhead{$\delta(J2000)$} &
\colhead{} &
\colhead{} &
\colhead{$(\log M_{\odot})$} &
\colhead{} &
\colhead{} &
\colhead{(Gyr)} &
\colhead{} &
\colhead{} &
\colhead{} &
\colhead{(mag)}
}

\startdata
1166 & ... & $9.3552761078$ & $0.9739487767$ & $0.38240 \pm 0.00050$ & $11.08$ & $11.15$ & $11.22$ & $4.13$ & $6.47$ & $7.53$ & $0.023 \pm 0.068$ & $1.274 \pm 1.103$ & $-0.3288 \pm 0.2179$ \\ 
1253 & 2005fd & $323.7985839844$ & $0.1628694236$ & $0.26200 \pm 0.00500$ & $11.27$ & $11.34$ & $11.39$ & $5.89$ & $7.80$ & $8.03$ & $-0.119 \pm 0.058$ & $-1.280 \pm 0.464$ & $-0.1097 \pm 0.1611$ \\ 
1371 & 2005fh & $349.3737487793$ & $0.4296737611$ & $0.11915 \pm 0.00012$ & $10.95$ & $11.00$ & $11.02$ & $5.27$ & $6.76$ & $7.26$ & $-0.084 \pm 0.020$ & $0.703 \pm 0.167$ & $-0.1775 \pm 0.0566$ \\ 
1580 & 2005fb & $45.3238296509$ & $-0.6422790885$ & $0.18300 \pm 0.00008$ & $10.61$ & $10.73$ & $10.83$ & $3.58$ & $5.20$ & $6.95$ & $-0.058 \pm 0.026$ & $0.675 \pm 0.271$ & $-0.1156 \pm 0.0775$ \\ 
1688 & ... & $321.3578186035$ & $0.3248503506$ & $0.35870 \pm 0.00050$ & $10.09$ & $10.20$ & $10.32$ & $1.50$ & $2.04$ & $2.74$ & $0.007 \pm 0.070$ & $1.019 \pm 1.306$ & $0.0240 \pm 0.2309$ \\ 
2017 & 2005fo & $328.9438781738$ & $0.5934827328$ & $0.26160 \pm 0.00050$ & $10.48$ & $10.54$ & $10.57$ & $4.26$ & $5.84$ & $6.66$ & $-0.117 \pm 0.052$ & $1.272 \pm 0.527$ & $0.2671 \pm 0.1512$ \\ 
2165 & 2005fr & $17.0916309357$ & $-0.0962756798$ & $0.28800 \pm 0.00500$ & $9.33$ & $9.39$ & $9.46$ & $1.86$ & $2.21$ & $3.52$ & $-0.130 \pm 0.038$ & $0.620 \pm 0.526$ & $0.3219 \pm 0.1315$ \\ 
2330 & 2005fp & $6.8073453903$ & $1.1208769083$ & $0.21320 \pm 0.00050$ & $9.83$ & $9.88$ & $9.94$ & $2.80$ & $4.11$ & $6.16$ & $0.083 \pm 0.063$ & $-2.238 \pm 0.569$ & $0.2954 \pm 0.1620$ \\ 
2372 & 2005ft & $40.5208168030$ & $-0.5410116911$ & $0.18050 \pm 0.00050$ & $10.37$ & $10.45$ & $10.49$ & $6.08$ & $7.64$ & $8.81$ & $0.045 \pm 0.024$ & $-0.015 \pm 0.225$ & $-0.1379 \pm 0.0714$ \\ 
2422 & 2005fi & $1.9945372343$ & $0.6381285191$ & $0.26500 \pm 0.00500$ & $9.09$ & $9.15$ & $9.25$ & $2.03$ & $2.57$ & $3.29$ & $-0.184 \pm 0.035$ & $0.751 \pm 0.326$ & $0.2078 \pm 0.1181$ \\ 
2533 & 2005fs & $31.2206439972$ & $-0.3263290226$ & $0.34000 \pm 0.00500$ & $9.95$ & $10.04$ & $10.14$ & $2.16$ & $3.94$ & $5.59$ & $-0.075 \pm 0.060$ & $2.329 \pm 0.727$ & $0.1465 \pm 0.1912$ \\ 
2635 & 2005fw & $52.7040061951$ & $-1.2376136780$ & $0.14370 \pm 0.00050$ & $9.93$ & $9.99$ & $10.04$ & $3.35$ & $4.73$ & $6.00$ & $-0.062 \pm 0.021$ & $0.839 \pm 0.183$ & $0.0388 \pm 0.0545$ \\ 
2789 & 2005fx & $344.2020263672$ & $0.4005828500$ & $0.29030 \pm 0.00050$ & $11.07$ & $11.15$ & $11.20$ & $5.35$ & $7.35$ & $9.35$ & $-0.115 \pm 0.051$ & $-0.292 \pm 0.543$ & $0.1218 \pm 0.1512$ \\ 
2943 & 2005go & $17.7050647736$ & $1.0080429316$ & $0.26540 \pm 0.00050$ & $9.00$ & $9.08$ & $9.17$ & $1.99$ & $2.53$ & $3.29$ & $-0.025 \pm 0.045$ & $0.121 \pm 0.408$ & $-0.0989 \pm 0.1285$ \\ 
3080 & 2005ga & $16.9316864014$ & $-1.0394667387$ & $0.17500 \pm 0.00050$ & $10.92$ & $10.97$ & $10.97$ & $3.67$ & $4.87$ & $4.87$ & $-0.068 \pm 0.025$ & $-0.166 \pm 0.244$ & $-0.1118 \pm 0.0621$ \\ 
3199 & 2005gs & $333.2925415039$ & $1.0506948233$ & $0.25110 \pm 0.00050$ & $8.79$ & $8.88$ & $8.98$ & $2.07$ & $2.66$ & $3.80$ & $-0.020 \pm 0.039$ & $1.166 \pm 0.414$ & $0.3073 \pm 0.1018$ \\ 
3256 & 2005hn & $329.2674865723$ & $-0.2234567255$ & $0.10760 \pm 0.00050$ & $9.76$ & $9.81$ & $9.84$ & $4.25$ & $4.78$ & $5.87$ & $0.034 \pm 0.034$ & $-0.714 \pm 0.198$ & $0.0294 \pm 0.0627$ \\ 
3377 & 2005gr & $54.1561660767$ & $1.0789009333$ & $0.24510 \pm 0.00050$ & $9.32$ & $9.38$ & $9.45$ & $2.15$ & $2.69$ & $3.84$ & $-0.100 \pm 0.034$ & $0.756 \pm 0.337$ & $-0.2046 \pm 0.0875$ \\ 
3451 & 2005gf & $334.0685424805$ & $0.7077997923$ & $0.25000 \pm 0.00050$ & $10.72$ & $10.81$ & $10.85$ & $4.74$ & $8.34$ & $9.34$ & $-0.073 \pm 0.035$ & $0.052 \pm 0.362$ & $-0.2537 \pm 0.0948$ \\ 
3452 & 2005gg & $334.6713256836$ & $0.6394435167$ & $0.23040 \pm 0.00050$ & $9.46$ & $9.47$ & $9.50$ & $2.15$ & $2.19$ & $2.73$ & $-0.100 \pm 0.034$ & $0.680 \pm 0.338$ & $-0.0220 \pm 0.0856$ \\ 
3592 & 2005gb & $19.0529479980$ & $0.7905687690$ & $0.08656 \pm 0.00019$ & $10.60$ & $10.64$ & $10.69$ & $6.16$ & $7.45$ & $8.40$ & $-0.031 \pm 0.019$ & $-0.454 \pm 0.146$ & $-0.0411 \pm 0.0513$ \\ 
4000 & 2005gt & $31.0166950226$ & $-0.3663079143$ & $0.27860 \pm 0.00050$ & $10.89$ & $10.95$ & $11.00$ & $4.39$ & $6.08$ & $7.37$ & $-0.032 \pm 0.074$ & $-0.990 \pm 0.619$ & $0.2036 \pm 0.2011$ \\ 
4046 & 2005gw & $354.4983215332$ & $0.6421458125$ & $0.27700 \pm 0.00500$ & $9.15$ & $9.27$ & $9.39$ & $2.51$ & $4.15$ & $6.01$ & $-0.026 \pm 0.041$ & $0.609 \pm 0.533$ & $0.0048 \pm 0.1390$ \\ 
4241 & 2005gu & $12.2376222610$ & $-0.9054884911$ & $0.33200 \pm 0.00050$ & $9.19$ & $9.25$ & $9.33$ & $1.61$ & $1.73$ & $2.26$ & $-0.068 \pm 0.053$ & $0.044 \pm 0.602$ & $0.0697 \pm 0.1566$ \\ 
4577 & 2005gv & $38.4758186340$ & $0.2808535695$ & $0.36300 \pm 0.00500$ & $10.54$ & $10.73$ & $10.83$ & $3.25$ & $4.52$ & $6.18$ & $-0.035 \pm 0.054$ & $0.178 \pm 0.691$ & $-0.1117 \pm 0.1733$ \\ 
4679 & 2005gy & $21.5282917023$ & $0.6768267751$ & $0.33240 \pm 0.00050$ & $9.43$ & $9.52$ & $9.60$ & $2.26$ & $3.05$ & $4.21$ & $0.075 \pm 0.058$ & $0.726 \pm 0.704$ & $-0.1301 \pm 0.1685$ \\ 
5103 & 2005gx & $359.8843383789$ & $0.7369195819$ & $0.16190 \pm 0.00050$ & $9.28$ & $9.28$ & $9.30$ & $2.57$ & $2.57$ & $2.62$ & $0.041 \pm 0.026$ & $-0.398 \pm 0.224$ & $-0.0962 \pm 0.0622$ \\ 
5183 & 2005gq & $53.4536514282$ & $0.7093452215$ & $0.38980 \pm 0.00050$ & $9.79$ & $9.90$ & $10.03$ & $1.47$ & $2.57$ & $4.37$ & $-0.139 \pm 0.077$ & $0.282 \pm 0.925$ & $0.1701 \pm 0.2238$ \\ 
5391 & 2005hs & $52.3423271179$ & $-1.0952030420$ & $0.30090 \pm 0.00050$ & $-99.00$ & $9.30$ & $9.36$ & $1.77$ & $1.79$ & $1.83$ & $-0.038 \pm 0.057$ & $0.023 \pm 0.581$ & $-0.0537 \pm 0.1579$ \\ 
5533 & 2005hu & $328.6699523926$ & $0.4132809639$ & $0.21970 \pm 0.00050$ & $9.69$ & $9.74$ & $9.79$ & $2.25$ & $2.79$ & $3.83$ & $0.048 \pm 0.021$ & $-0.035 \pm 0.336$ & $-0.0135 \pm 0.0713$ \\ 
5736 & 2005jz & $22.8627357483$ & $-0.6316036582$ & $0.25300 \pm 0.00500$ & $8.89$ & $8.97$ & $9.06$ & $2.06$ & $2.30$ & $3.25$ & $-0.009 \pm 0.025$ & $-0.530 \pm 0.335$ & $-0.0444 \pm 0.1002$ \\ 
5737 & 2005ib & $22.8571491241$ & $-0.6033283472$ & $0.39300 \pm 0.00050$ & $9.71$ & $9.80$ & $9.90$ & $1.90$ & $2.49$ & $3.63$ & $0.079 \pm 0.066$ & $1.329 \pm 1.078$ & $-0.2717 \pm 0.1697$ \\ 
5844 & 2005ic & $327.7861633301$ & $-0.8428391814$ & $0.31080 \pm 0.00050$ & $9.41$ & $9.48$ & $9.56$ & $1.72$ & $1.85$ & $2.50$ & $-0.115 \pm 0.035$ & $-0.087 \pm 0.495$ & $-0.0486 \pm 0.0965$ \\ 
5944 & 2005hc & $29.2021064758$ & $-0.2125778049$ & $0.04594 \pm 0.00017$ & $10.87$ & $10.87$ & $10.92$ & $7.26$ & $7.26$ & $10.33$ & $-0.025 \pm 0.015$ & $0.543 \pm 0.127$ & $0.0667 \pm 0.0739$ \\ 
5957 & 2005ie & $34.7598075867$ & $-0.2725664973$ & $0.27960 \pm 0.00050$ & $10.45$ & $10.49$ & $10.52$ & $1.91$ & $2.45$ & $4.25$ & $-0.104 \pm 0.038$ & $-0.771 \pm 0.489$ & $-0.0143 \pm 0.1051$ \\ 
6100 & 2005ka & $333.4833679199$ & $1.0861500502$ & $0.31770 \pm 0.00050$ & $9.99$ & $10.00$ & $10.05$ & $1.68$ & $1.74$ & $2.24$ & $0.027 \pm 0.067$ & $2.730 \pm 1.223$ & $0.4055 \pm 0.2179$ \\ 
6196 & 2005ig & $337.6311950684$ & $-0.5023938417$ & $0.28070 \pm 0.00050$ & $11.11$ & $11.17$ & $11.23$ & $6.44$ & $8.44$ & $9.54$ & $0.006 \pm 0.046$ & $-0.899 \pm 0.653$ & $-0.2086 \pm 0.1329$ \\ 
6649 & 2005jd & $34.2763099670$ & $0.5356944203$ & $0.31400 \pm 0.00500$ & $9.34$ & $9.45$ & $9.54$ & $4.12$ & $5.75$ & $6.50$ & $-0.102 \pm 0.042$ & $0.568 \pm 0.606$ & $-0.0221 \pm 0.1289$ \\ 
6777 & 2005iy & $321.2164306641$ & $0.3856396675$ & $0.40430 \pm 0.00050$ & $9.11$ & $9.26$ & $9.44$ & $1.35$ & $2.39$ & $3.79$ & $-0.039 \pm 0.089$ & $1.716 \pm 1.330$ & $0.2246 \pm 0.2238$ \\ 
6936 & 2005jl & $323.2338867188$ & $-0.7000573874$ & $0.18100 \pm 0.00050$ & $10.18$ & $10.26$ & $10.33$ & $3.40$ & $4.90$ & $6.68$ & $-0.011 \pm 0.027$ & $-0.144 \pm 0.315$ & $-0.0136 \pm 0.0735$ \\ 
7143 & 2005jg & $345.2623596191$ & $-0.2068603635$ & $0.30400 \pm 0.00500$ & $10.30$ & $10.36$ & $10.40$ & $3.49$ & $5.09$ & $6.24$ & $0.033 \pm 0.040$ & $-0.894 \pm 0.590$ & $-0.0097 \pm 0.1264$ \\ 
7512 & 2005jo & $52.0903663635$ & $-0.3261369467$ & $0.21900 \pm 0.00500$ & $-99.00$ & $8.62$ & $8.67$ & $2.22$ & $2.26$ & $2.43$ & $0.013 \pm 0.031$ & $0.279 \pm 0.425$ & $0.0395 \pm 0.1191$ \\ 
7847 & 2005jp & $32.4597015381$ & $-0.0616886541$ & $0.21240 \pm 0.00050$ & $10.43$ & $10.49$ & $10.58$ & $3.28$ & $4.25$ & $7.31$ & $0.174 \pm 0.030$ & $0.255 \pm 0.410$ & $-0.1968 \pm 0.0875$ \\ 
8030 & 2005jv & $40.2087211609$ & $0.9932332635$ & $0.42200 \pm 0.00500$ & $9.41$ & $9.56$ & $9.72$ & $1.30$ & $2.35$ & $3.95$ & $-0.165 \pm 0.082$ & $0.923 \pm 1.127$ & $0.3907 \pm 0.2467$ \\ 
8213 & 2005ko & $357.5210571289$ & $-0.9214569926$ & $0.18470 \pm 0.00050$ & $10.38$ & $10.45$ & $10.49$ & $4.35$ & $5.97$ & $7.09$ & $0.216 \pm 0.040$ & $-0.651 \pm 0.377$ & $-0.1624 \pm 0.0913$ \\ 
8598 & 2005jt & $42.6674690247$ & $-0.0667039678$ & $0.36060 \pm 0.00050$ & $10.01$ & $10.11$ & $10.19$ & $1.63$ & $2.11$ & $2.73$ & $0.090 \pm 0.085$ & $-1.283 \pm 0.940$ & $-0.3454 \pm 0.2274$ \\ 
8719 & 2005kp & $7.7218842506$ & $-0.7186533809$ & $0.11780 \pm 0.00050$ & $9.01$ & $9.10$ & $9.14$ & $2.92$ & $3.46$ & $3.54$ & $-0.099 \pm 0.026$ & $-0.263 \pm 0.259$ & $0.0664 \pm 0.0649$ \\ 
9207 & 2005lg & $19.0833320618$ & $-0.8073787689$ & $0.35000 \pm 0.00050$ & $10.73$ & $10.79$ & $10.85$ & $4.10$ & $5.26$ & $5.99$ & $0.002 \pm 0.066$ & $1.512 \pm 0.916$ & $0.0289 \pm 0.1974$ \\ 
9457 & 2005li & $335.8146362305$ & $0.2536900640$ & $0.25690 \pm 0.00050$ & $11.00$ & $11.05$ & $11.10$ & $7.67$ & $8.77$ & $10.27$ & $0.001 \pm 0.065$ & $-0.186 \pm 0.819$ & $-0.0116 \pm 0.1659$ \\ 
10550 & 2005lf & $349.6758117676$ & $-1.2046753168$ & $0.30010 \pm 0.00050$ & $10.33$ & $10.38$ & $10.39$ & $1.77$ & $1.80$ & $3.26$ & $0.068 \pm 0.075$ & $1.281 \pm 1.130$ & $0.1359 \pm 0.2309$ \\ 
12781 & 2006er & $5.4078617096$ & $-1.0106090307$ & $0.08431 \pm 0.00016$ & $10.96$ & $10.97$ & $11.02$ & $10.58$ & $11.18$ & $11.58$ & $0.072 \pm 0.061$ & $-2.128 \pm 0.337$ & $0.1604 \pm 0.1149$ \\ 
12843 & 2006fa & $323.8784790039$ & $-0.9796369672$ & $0.16704 \pm 0.00013$ & $11.18$ & $11.22$ & $11.28$ & $7.61$ & $8.71$ & $10.61$ & $0.082 \pm 0.043$ & $-1.110 \pm 0.405$ & $-0.3135 \pm 0.1016$ \\ 
12856 & 2006fl & $332.8653564453$ & $0.7555990219$ & $0.17173 \pm 0.00011$ & $10.32$ & $10.37$ & $10.45$ & $3.16$ & $3.93$ & $6.07$ & $-0.130 \pm 0.023$ & $0.738 \pm 0.310$ & $0.0908 \pm 0.0711$ \\ 
12860 & 2006fc & $323.6949768066$ & $1.1754231453$ & $0.12170 \pm 0.00050$ & $10.58$ & $10.63$ & $10.67$ & $3.43$ & $5.02$ & $6.17$ & $0.186 \pm 0.024$ & $-0.494 \pm 0.258$ & $-0.0007 \pm 0.0649$ \\ 
12898 & 2006fw & $26.7930507660$ & $-0.1468682140$ & $0.08350 \pm 0.00050$ & $9.96$ & $9.97$ & $10.01$ & $6.93$ & $7.65$ & $8.43$ & $0.074 \pm 0.019$ & $-0.332 \pm 0.139$ & $0.0141 \pm 0.0583$ \\ 
12930 & 2006ex & $309.6826477051$ & $-0.4763843715$ & $0.14749 \pm 0.00017$ & $10.85$ & $10.88$ & $10.90$ & $4.12$ & $4.61$ & $5.12$ & $-0.023 \pm 0.037$ & $1.623 \pm 0.477$ & $0.1716 \pm 0.0947$ \\ 
12950 & 2006fy & $351.6672668457$ & $-0.8406041265$ & $0.08268 \pm 0.00004$ & $9.75$ & $9.78$ & $9.81$ & $3.18$ & $3.81$ & $4.34$ & $0.028 \pm 0.014$ & $-0.731 \pm 0.112$ & $0.0301 \pm 0.0480$ \\ 
12972 & 2006ft & $7.9585695267$ & $-0.3830518126$ & $0.26080 \pm 0.00050$ & $9.12$ & $9.18$ & $9.27$ & $2.06$ & $2.67$ & $3.83$ & $-0.059 \pm 0.044$ & $0.723 \pm 0.696$ & $0.1218 \pm 0.1387$ \\ 
13044 & 2006fm & $332.5429992676$ & $0.5039222836$ & $0.12570 \pm 0.00050$ & $9.66$ & $9.70$ & $9.76$ & $3.30$ & $4.12$ & $5.72$ & $-0.083 \pm 0.021$ & $-0.197 \pm 0.204$ & $0.0565 \pm 0.0547$ \\ 
13070 & 2006fu & $357.7849121094$ & $-0.7465677261$ & $0.19855 \pm 0.00009$ & $10.18$ & $10.22$ & $10.23$ & $3.08$ & $3.74$ & $4.13$ & $-0.179 \pm 0.027$ & $0.717 \pm 0.323$ & $0.0957 \pm 0.0754$ \\ 
13305 & 2006he & $331.1001586914$ & $0.6907849908$ & $0.21390 \pm 0.00050$ & $10.01$ & $10.06$ & $10.11$ & $2.58$ & $3.50$ & $4.63$ & $-0.022 \pm 0.030$ & $1.020 \pm 0.352$ & $0.0304 \pm 0.0817$ \\ 
13354 & 2006hr & $27.5647277832$ & $-0.8866921663$ & $0.15760 \pm 0.00010$ & $10.46$ & $10.51$ & $10.56$ & $3.59$ & $4.73$ & $6.74$ & $0.087 \pm 0.024$ & $0.952 \pm 0.225$ & $-0.0979 \pm 0.0592$ \\ 
13411 & ... & $315.1897277832$ & $0.1917154342$ & $0.16300 \pm 0.00050$ & $9.13$ & $9.21$ & $9.29$ & $3.74$ & $5.40$ & $6.90$ & $-0.026 \pm 0.034$ & $1.191 \pm 0.397$ & $0.1275 \pm 0.0895$ \\ 
13425 & 2006gp & $338.5414733887$ & $0.0548623651$ & $0.21290 \pm 0.00050$ & $10.22$ & $10.29$ & $10.36$ & $5.72$ & $8.28$ & $10.19$ & $0.298 \pm 0.054$ & $-0.751 \pm 0.579$ & $-0.0301 \pm 0.1606$ \\ 
13506 & 2006hg & $25.2436542511$ & $-0.7284323573$ & $0.24500 \pm 0.00050$ & $10.13$ & $10.18$ & $10.22$ & $2.35$ & $3.93$ & $5.25$ & $0.165 \pm 0.038$ & $0.079 \pm 0.582$ & $0.0435 \pm 0.1271$ \\ 
13511 & 2006hh & $40.6112861633$ & $-0.7942346931$ & $0.23757 \pm 0.00015$ & $11.26$ & $11.41$ & $11.47$ & $4.98$ & $7.34$ & $8.03$ & $-0.092 \pm 0.044$ & $-1.991 \pm 0.438$ & $0.1846 \pm 0.1132$ \\ 
13578 & 2006hc & $17.3948116302$ & $0.7042742372$ & $0.22900 \pm 0.00050$ & $9.11$ & $9.19$ & $9.29$ & $2.36$ & $3.33$ & $4.79$ & $-0.043 \pm 0.029$ & $0.150 \pm 0.464$ & $0.1010 \pm 0.1018$ \\ 
13641 & 2006hf & $345.2174987793$ & $-0.9820173383$ & $0.21930 \pm 0.00050$ & $9.15$ & $9.25$ & $9.34$ & $2.79$ & $4.10$ & $6.12$ & $-0.046 \pm 0.029$ & $0.967 \pm 0.322$ & $-0.0431 \pm 0.0777$ \\ 
13736 & 2006hv & $336.8327026367$ & $1.0307192802$ & $0.15040 \pm 0.00050$ & $9.51$ & $9.56$ & $9.61$ & $2.93$ & $3.84$ & $5.51$ & $-0.040 \pm 0.023$ & $0.947 \pm 0.232$ & $0.0361 \pm 0.0597$ \\ 
13757 & 2006hk & $350.1237792969$ & $-1.1580305099$ & $0.28900 \pm 0.00500$ & $9.13$ & $9.24$ & $9.36$ & $1.98$ & $2.98$ & $4.49$ & $-0.203 \pm 0.039$ & $0.724 \pm 0.416$ & $0.1401 \pm 0.1212$ \\ 
13796 & 2006hl & $350.6919860840$ & $0.5323168635$ & $0.14820 \pm 0.00050$ & $10.16$ & $10.22$ & $10.27$ & $4.41$ & $5.99$ & $7.12$ & $-0.054 \pm 0.021$ & $0.518 \pm 0.181$ & $-0.0995 \pm 0.0545$ \\ 
13835 & 2006hp & $6.0593752861$ & $-0.2492461652$ & $0.24770 \pm 0.00050$ & $10.39$ & $10.44$ & $10.48$ & $2.24$ & $3.22$ & $3.92$ & $-0.064 \pm 0.031$ & $0.707 \pm 0.295$ & $0.0003 \pm 0.0836$ \\ 
13894 & 2006jh & $1.6905879974$ & $-0.0367476977$ & $0.12490 \pm 0.00050$ & $9.34$ & $9.39$ & $9.44$ & $4.09$ & $6.15$ & $7.58$ & $0.186 \pm 0.022$ & $0.446 \pm 0.201$ & $0.1657 \pm 0.0547$ \\ 
13934 & 2006jg & $342.1104431152$ & $-0.4351437390$ & $0.33000 \pm 0.00500$ & $10.76$ & $10.83$ & $10.90$ & $2.69$ & $4.58$ & $6.13$ & $-0.079 \pm 0.078$ & $-0.914 \pm 0.786$ & $0.2149 \pm 0.2178$ \\ 
13956 & 2006hi & $20.9414615631$ & $0.8162868619$ & $0.26200 \pm 0.00500$ & $10.53$ & $10.61$ & $10.77$ & $6.22$ & $8.62$ & $9.72$ & $0.144 \pm 0.069$ & $-0.552 \pm 1.204$ & $0.6144 \pm 0.2207$ \\ 
14019 & 2006ki & $316.6423950195$ & $-0.6486185193$ & $0.21640 \pm 0.00050$ & $9.69$ & $9.74$ & $9.82$ & $2.33$ & $2.96$ & $4.12$ & $0.001 \pm 0.041$ & $-0.337 \pm 0.389$ & $0.3925 \pm 0.1018$ \\ 
14108 & 2006hu & $53.5947074890$ & $-1.1231447458$ & $0.13300 \pm 0.00500$ & $8.77$ & $8.86$ & $8.96$ & $3.53$ & $4.83$ & $6.79$ & $-0.001 \pm 0.020$ & $0.053 \pm 0.164$ & $0.2947 \pm 0.1352$ \\ 
14212 & 2006iy & $330.4706420898$ & $1.0444601774$ & $0.20540 \pm 0.00050$ & $10.27$ & $10.35$ & $10.42$ & $4.58$ & $6.55$ & $8.01$ & $-0.009 \pm 0.023$ & $-0.415 \pm 0.205$ & $0.0090 \pm 0.0620$ \\ 
14261 & 2006jk & $328.2404174805$ & $0.2536858320$ & $0.28580 \pm 0.00050$ & $9.39$ & $9.44$ & $9.52$ & $1.85$ & $1.91$ & $2.52$ & $-0.037 \pm 0.039$ & $0.523 \pm 0.543$ & $-0.1989 \pm 0.1132$ \\ 
14298 & 2006jj & $314.8951110840$ & $1.2232679129$ & $0.27010 \pm 0.00050$ & $9.28$ & $9.41$ & $9.54$ & $2.50$ & $3.92$ & $5.77$ & $-0.085 \pm 0.039$ & $1.049 \pm 0.442$ & $0.0262 \pm 0.1084$ \\ 
14331 & 2006kl & $7.8891010284$ & $-0.1355372667$ & $0.22110 \pm 0.00050$ & $9.57$ & $9.61$ & $9.66$ & $2.30$ & $2.78$ & $3.45$ & $0.002 \pm 0.033$ & $-0.140 \pm 0.313$ & $0.1347 \pm 0.0875$ \\ 
14397 & 2006kk & $6.9156045914$ & $0.6493207216$ & $0.38570 \pm 0.00050$ & $10.55$ & $10.61$ & $10.71$ & $1.98$ & $2.59$ & $4.17$ & $-0.321 \pm 0.083$ & $-1.077 \pm 0.745$ & $0.4810 \pm 0.2402$ \\ 
14437 & 2006hy & $332.0809326172$ & $-1.1963416338$ & $0.14910 \pm 0.00050$ & $9.96$ & $10.02$ & $10.09$ & $5.81$ & $8.81$ & $10.41$ & $-0.106 \pm 0.021$ & $0.278 \pm 0.192$ & $-0.0850 \pm 0.0544$ \\ 
14456 & 2006jm & $343.5509338379$ & $1.0508996248$ & $0.33000 \pm 0.00500$ & $11.04$ & $11.13$ & $11.19$ & $4.39$ & $5.67$ & $6.97$ & $0.001 \pm 0.042$ & $0.163 \pm 0.549$ & $-0.2071 \pm 0.1369$ \\ 
14481 & 2006lj & $2.6814725399$ & $0.2018533349$ & $0.24390 \pm 0.00050$ & $10.96$ & $11.08$ & $11.17$ & $4.92$ & $8.95$ & $8.95$ & $-0.168 \pm 0.050$ & $-1.208 \pm 0.443$ & $0.3927 \pm 0.1344$ \\ 
14735 & 2006km & $35.1584739685$ & $0.3481049836$ & $0.30110 \pm 0.00050$ & $10.29$ & $10.38$ & $10.47$ & $2.91$ & $4.41$ & $5.72$ & $0.045 \pm 0.039$ & $0.157 \pm 0.407$ & $-0.2168 \pm 0.1067$ \\ 
14782 & 2006jp & $314.2340698242$ & $-0.2791627347$ & $0.16040 \pm 0.00050$ & $11.13$ & $11.26$ & $11.35$ & $4.80$ & $7.28$ & $9.68$ & $0.011 \pm 0.023$ & $-0.503 \pm 0.211$ & $-0.4487 \pm 0.0597$ \\ 
14815 & 2006iz & $319.0716552734$ & $0.5595042109$ & $0.13630 \pm 0.00050$ & $8.78$ & $8.79$ & $8.82$ & $2.74$ & $2.74$ & $2.79$ & $-0.027 \pm 0.042$ & $3.869 \pm 0.430$ & $0.2070 \pm 0.1037$ \\ 
14846 & 2006jn & $7.6626000404$ & $0.1420275271$ & $0.22470 \pm 0.00050$ & $10.96$ & $10.99$ & $11.04$ & $4.23$ & $5.68$ & $6.97$ & $-0.059 \pm 0.030$ & $0.378 \pm 0.333$ & $0.0229 \pm 0.0894$ \\ 
14871 & 2006jq & $54.2769241333$ & $0.0092711495$ & $0.12760 \pm 0.00050$ & $9.21$ & $9.26$ & $9.31$ & $2.85$ & $3.84$ & $5.27$ & $-0.086 \pm 0.019$ & $1.128 \pm 0.178$ & $0.0349 \pm 0.0518$ \\ 
14979 & 2006jr & $54.9465255737$ & $0.9921327233$ & $0.17710 \pm 0.00050$ & $10.00$ & $10.01$ & $10.10$ & $3.96$ & $4.42$ & $6.78$ & $-0.119 \pm 0.021$ & $-0.055 \pm 0.192$ & $-0.0599 \pm 0.0596$ \\ 
15132 & 2006jt & $329.6999511719$ & $0.1987692863$ & $0.14400 \pm 0.00500$ & $9.45$ & $9.46$ & $9.47$ & $2.69$ & $2.69$ & $2.74$ & $-0.138 \pm 0.021$ & $0.752 \pm 0.211$ & $0.1824 \pm 0.1300$ \\ 
15201 & 2006ks & $337.5189208984$ & $0.0031410647$ & $0.20850 \pm 0.00050$ & $11.26$ & $11.34$ & $11.41$ & $6.16$ & $8.16$ & $10.26$ & $0.129 \pm 0.038$ & $-1.386 \pm 0.474$ & $0.3592 \pm 0.1117$ \\ 
15203 & 2006jy & $15.7347574234$ & $0.1830275059$ & $0.20430 \pm 0.00050$ & $10.14$ & $10.22$ & $10.26$ & $5.21$ & $8.21$ & $9.81$ & $0.001 \pm 0.028$ & $1.139 \pm 0.335$ & $-0.0040 \pm 0.0797$ \\ 
15213 & 2006lk & $53.0192298889$ & $-0.1002237424$ & $0.31120 \pm 0.00050$ & $10.58$ & $10.62$ & $10.67$ & $4.67$ & $5.82$ & $6.52$ & $-0.042 \pm 0.051$ & $-0.247 \pm 0.570$ & $-0.1117 \pm 0.1401$ \\ 
15217 & 2006jv & $22.6341056824$ & $0.2209988385$ & $0.36800 \pm 0.00500$ & $10.71$ & $10.79$ & $10.87$ & $3.80$ & $4.77$ & $6.65$ & $-0.056 \pm 0.116$ & $-1.623 \pm 1.142$ & $-0.0224 \pm 0.3205$ \\ 
15219 & 2006ka & $34.6107521057$ & $0.2261287570$ & $0.24800 \pm 0.00500$ & $10.94$ & $11.02$ & $11.08$ & $2.51$ & $3.62$ & $5.48$ & $-0.165 \pm 0.034$ & $-0.294 \pm 0.522$ & $0.1063 \pm 0.1252$ \\ 
15229 & 2006kr & $4.8320274353$ & $1.0906258821$ & $0.22680 \pm 0.00050$ & $9.09$ & $9.12$ & $9.18$ & $2.17$ & $2.27$ & $2.89$ & $-0.048 \pm 0.032$ & $0.575 \pm 0.362$ & $0.0729 \pm 0.0912$ \\ 
15259 & 2006kc & $337.5441894531$ & $-0.4077875614$ & $0.21003 \pm 0.00011$ & $9.14$ & $9.25$ & $9.36$ & $3.00$ & $4.82$ & $6.66$ & $-0.005 \pm 0.027$ & $0.184 \pm 0.290$ & $-0.1254 \pm 0.0775$ \\ 
15287 & 2006kt & $323.9606628418$ & $-1.0589238405$ & $0.25400 \pm 0.00500$ & $10.47$ & $10.58$ & $10.65$ & $4.30$ & $6.70$ & $8.85$ & $-0.080 \pm 0.027$ & $0.859 \pm 0.350$ & $0.0237 \pm 0.1050$ \\ 
15354 & 2006lp & $6.7742686272$ & $-0.1259586960$ & $0.22210 \pm 0.00050$ & $10.82$ & $10.85$ & $10.88$ & $9.02$ & $10.12$ & $11.02$ & $0.138 \pm 0.056$ & $-2.119 \pm 0.458$ & $-0.1773 \pm 0.1401$ \\ 
15356 & 2006lm & $335.0533142090$ & $0.4099416137$ & $0.27470 \pm 0.00050$ & $10.41$ & $10.54$ & $10.62$ & $4.62$ & $6.72$ & $8.65$ & $-0.018 \pm 0.046$ & $-0.566 \pm 0.530$ & $-0.1923 \pm 0.1300$ \\ 
15369 & 2006ln & $348.8330383301$ & $-0.5626841784$ & $0.23200 \pm 0.00500$ & $8.95$ & $9.06$ & $9.18$ & $2.72$ & $4.02$ & $5.83$ & $-0.014 \pm 0.026$ & $0.624 \pm 0.304$ & $-0.4142 \pm 0.1023$ \\ 
15383 & 2006lq & $34.1496849060$ & $-0.1552789956$ & $0.31620 \pm 0.00050$ & $10.54$ & $10.67$ & $10.77$ & $3.71$ & $5.22$ & $8.11$ & $-0.042 \pm 0.049$ & $-0.332 \pm 0.622$ & $-0.0459 \pm 0.1512$ \\ 
15421 & 2006kw & $33.7412719727$ & $0.6027206182$ & $0.18500 \pm 0.00050$ & $10.09$ & $10.14$ & $10.22$ & $3.01$ & $4.10$ & $5.68$ & $-0.031 \pm 0.024$ & $-0.104 \pm 0.297$ & $0.0667 \pm 0.0691$ \\ 
15425 & 2006kx & $55.5610733032$ & $0.4783548415$ & $0.16004 \pm 0.00014$ & $10.50$ & $10.57$ & $10.59$ & $4.69$ & $6.29$ & $8.69$ & $-0.031 \pm 0.021$ & $0.834 \pm 0.257$ & $-0.2347 \pm 0.0592$ \\ 
15440 & 2006lr & $39.7205924988$ & $0.0901087895$ & $0.26190 \pm 0.00050$ & $10.59$ & $10.71$ & $10.83$ & $3.76$ & $6.23$ & $7.80$ & $0.097 \pm 0.038$ & $-0.767 \pm 0.546$ & $-0.2466 \pm 0.1116$ \\ 
15443 & 2006lb & $49.8674354553$ & $-0.3179923296$ & $0.18202 \pm 0.00010$ & $10.26$ & $10.32$ & $10.37$ & $4.38$ & $6.51$ & $7.36$ & $-0.077 \pm 0.025$ & $1.357 \pm 0.304$ & $-0.0756 \pm 0.0592$ \\ 
15453 & 2006ky & $319.6684570312$ & $-1.0242410898$ & $0.18370 \pm 0.00050$ & $8.95$ & $9.04$ & $9.14$ & $3.08$ & $4.53$ & $6.17$ & $-0.021 \pm 0.031$ & $1.279 \pm 0.390$ & $0.1745 \pm 0.0757$ \\ 
15456 & 2006ll & $331.8668518066$ & $-0.9038056135$ & $0.38210 \pm 0.00050$ & $10.93$ & $10.97$ & $11.02$ & $5.08$ & $5.76$ & $6.15$ & $-0.128 \pm 0.083$ & $-0.712 \pm 1.053$ & $-0.1778 \pm 0.2437$ \\ 
15459 & 2006la & $340.7014160156$ & $-0.9017586112$ & $0.12670 \pm 0.00050$ & $9.04$ & $9.04$ & $9.06$ & $2.81$ & $2.81$ & $2.86$ & $0.138 \pm 0.026$ & $0.225 \pm 0.257$ & $0.4487 \pm 0.0599$ \\ 
15461 & 2006kz & $326.8482055664$ & $-0.4947634041$ & $0.18000 \pm 0.00500$ & $10.30$ & $10.35$ & $10.42$ & $4.82$ & $6.09$ & $8.25$ & $-0.083 \pm 0.027$ & $-0.371 \pm 0.280$ & $-0.0081 \pm 0.1148$ \\ 
15466 & 2006mz & $317.6454467773$ & $-0.1227247939$ & $0.24610 \pm 0.00050$ & $10.50$ & $10.56$ & $10.61$ & $2.53$ & $3.93$ & $5.74$ & $0.070 \pm 0.051$ & $-1.494 \pm 0.462$ & $-0.3856 \pm 0.1195$ \\ 
15467 & ... & $320.0201110840$ & $-0.1773548573$ & $0.21043 \pm 0.00009$ & $10.35$ & $10.35$ & $10.37$ & $2.27$ & $2.31$ & $2.85$ & $-0.042 \pm 0.032$ & $0.848 \pm 0.391$ & $-0.1144 \pm 0.0815$ \\ 
15504 & 2006oc & $345.7013854980$ & $-0.8768699169$ & $0.27010 \pm 0.00050$ & $11.03$ & $11.08$ & $11.14$ & $3.15$ & $4.20$ & $6.06$ & $0.257 \pm 0.061$ & $3.638 \pm 0.866$ & $0.0931 \pm 0.1800$ \\ 
15508 & 2006ls & $27.1694507599$ & $-0.5757497549$ & $0.14740 \pm 0.00050$ & $9.85$ & $9.88$ & $9.91$ & $2.80$ & $3.39$ & $3.96$ & $-0.084 \pm 0.021$ & $0.574 \pm 0.217$ & $-0.1928 \pm 0.0516$ \\ 
15583 & 2006mv & $37.7310752869$ & $0.9462816715$ & $0.17520 \pm 0.00050$ & $9.08$ & $9.13$ & $9.19$ & $2.74$ & $3.24$ & $4.44$ & $0.069 \pm 0.029$ & $-0.474 \pm 0.282$ & $-0.0197 \pm 0.0714$ \\ 
15648 & 2006ni & $313.7187805176$ & $-0.1958119273$ & $0.17496 \pm 0.00017$ & $11.23$ & $11.29$ & $11.36$ & $6.12$ & $8.62$ & $10.62$ & $0.186 \pm 0.049$ & $-1.277 \pm 0.511$ & $-0.0871 \pm 0.1083$ \\ 
15704 & 2006nh & $40.2121353149$ & $0.6598128676$ & $0.36500 \pm 0.00500$ & $10.78$ & $10.83$ & $10.88$ & $5.94$ & $6.89$ & $7.17$ & $-0.096 \pm 0.066$ & $0.706 \pm 0.874$ & $0.1095 \pm 0.1858$ \\ 
15776 & 2006na & $32.8302955627$ & $-0.9981175065$ & $0.30500 \pm 0.00500$ & $11.18$ & $11.19$ & $11.21$ & $9.81$ & $10.21$ & $10.21$ & $-0.116 \pm 0.081$ & $-1.662 \pm 0.743$ & $0.1115 \pm 0.2089$ \\ 
15872 & 2006nb & $36.7223777771$ & $-0.3278448582$ & $0.18460 \pm 0.00050$ & $9.52$ & $9.58$ & $9.64$ & $3.07$ & $4.35$ & $6.38$ & $-0.027 \pm 0.035$ & $0.791 \pm 0.460$ & $0.0956 \pm 0.0857$ \\ 
15897 & 2006pb & $11.6815948486$ & $-1.0324945450$ & $0.17470 \pm 0.00050$ & $10.70$ & $10.78$ & $10.87$ & $7.12$ & $8.12$ & $10.12$ & $0.073 \pm 0.052$ & $-2.887 \pm 0.338$ & $-0.1306 \pm 0.1019$ \\ 
15901 & 2006od & $31.9762687683$ & $-0.5353427529$ & $0.20530 \pm 0.00050$ & $9.84$ & $9.92$ & $10.00$ & $3.36$ & $4.70$ & $6.22$ & $-0.079 \pm 0.030$ & $-0.420 \pm 0.332$ & $0.0116 \pm 0.0756$ \\ 
16000 & 2006nj & $21.1174659729$ & $0.0743126571$ & $0.39000 \pm 0.00500$ & $9.26$ & $9.41$ & $9.57$ & $1.42$ & $2.29$ & $4.15$ & $-0.163 \pm 0.072$ & $1.448 \pm 1.091$ & $0.3161 \pm 0.1916$ \\ 
16072 & 2006nv & $3.1245520115$ & $-0.9778423309$ & $0.28670 \pm 0.00050$ & $10.80$ & $10.83$ & $10.96$ & $4.51$ & $5.49$ & $8.52$ & $-0.044 \pm 0.049$ & $0.058 \pm 0.691$ & $-0.0728 \pm 0.1285$ \\ 
16073 & 2006of & $8.1076574326$ & $-1.0539033413$ & $0.15310 \pm 0.00050$ & $9.68$ & $9.73$ & $9.78$ & $3.62$ & $5.08$ & $6.40$ & $-0.018 \pm 0.017$ & $0.080 \pm 0.252$ & $0.1098 \pm 0.0516$ \\ 
16099 & 2006nn & $26.4212512970$ & $-1.0545672178$ & $0.19686 \pm 0.00015$ & $10.49$ & $10.55$ & $10.61$ & $4.66$ & $6.29$ & $7.72$ & $0.004 \pm 0.025$ & $1.934 \pm 0.581$ & $0.1190 \pm 0.0855$ \\ 
16100 & 2006nl & $30.4363574982$ & $-1.0323493481$ & $0.19500 \pm 0.00500$ & $9.32$ & $9.42$ & $9.50$ & $4.90$ & $7.41$ & $9.90$ & $0.097 \pm 0.033$ & $-0.355 \pm 0.451$ & $0.0264 \pm 0.1237$ \\ 
16106 & 2006no & $332.0898742676$ & $-1.1483064890$ & $0.25120 \pm 0.00050$ & $10.80$ & $10.92$ & $10.98$ & $3.73$ & $4.33$ & $7.33$ & $-0.115 \pm 0.044$ & $-0.167 \pm 0.629$ & $-0.0047 \pm 0.1344$ \\ 
16185 & 2006ok & $16.8680858612$ & $-0.2693305314$ & $0.09700 \pm 0.00500$ & $9.59$ & $9.64$ & $9.69$ & $6.42$ & $8.58$ & $9.56$ & $0.178 \pm 0.031$ & $-1.614 \pm 0.277$ & $0.1008 \pm 0.1850$ \\ 
16232 & 2006oj & $17.2049808502$ & $-0.9894958138$ & $0.36700 \pm 0.00500$ & $10.54$ & $10.62$ & $10.70$ & $2.99$ & $5.00$ & $5.93$ & $-0.117 \pm 0.082$ & $-0.246 \pm 0.862$ & $0.0913 \pm 0.1821$ \\ 
17168 & 2007ik & $339.7236328125$ & $-1.1672555208$ & $0.18400 \pm 0.00500$ & $9.46$ & $9.50$ & $9.54$ & $2.54$ & $2.81$ & $3.61$ & $-0.016 \pm 0.034$ & $0.280 \pm 0.382$ & $0.0588 \pm 0.1338$ \\ 
17332 & 2007jk & $43.7725067139$ & $-0.1476856470$ & $0.18284 \pm 0.00015$ & $10.43$ & $10.55$ & $10.64$ & $3.91$ & $6.66$ & $8.68$ & $0.088 \pm 0.032$ & $-0.254 \pm 0.314$ & $-0.1682 \pm 0.0947$ \\ 
17366 & 2007hz & $315.7849731445$ & $-1.0311613083$ & $0.13933 \pm 0.00017$ & $10.87$ & $10.92$ & $10.97$ & $3.74$ & $5.18$ & $6.74$ & $-0.125 \pm 0.025$ & $0.588 \pm 0.253$ & $-0.0077 \pm 0.0689$ \\ 
17389 & 2007ih & $323.2950134277$ & $-0.9600833058$ & $0.17060 \pm 0.00050$ & $9.82$ & $9.90$ & $9.97$ & $3.68$ & $4.99$ & $6.84$ & $0.063 \pm 0.033$ & $1.104 \pm 0.406$ & $0.2739 \pm 0.1036$ \\ 
17497 & 2007jt & $37.1364936829$ & $-1.0428131819$ & $0.14478 \pm 0.00010$ & $10.33$ & $10.39$ & $10.43$ & $2.98$ & $3.91$ & $4.63$ & $0.057 \pm 0.020$ & $0.596 \pm 0.180$ & $-0.1025 \pm 0.0510$ \\ 
17552 & 2007jl & $322.3212585449$ & $-1.0028200150$ & $0.25420 \pm 0.00050$ & $10.55$ & $10.61$ & $10.66$ & $2.10$ & $3.36$ & $5.19$ & $-0.016 \pm 0.039$ & $0.766 \pm 0.403$ & $0.0018 \pm 0.1132$ \\ 
17568 & 2007kb & $313.1032714844$ & $0.2774721682$ & $0.14450 \pm 0.00050$ & $9.93$ & $10.00$ & $10.06$ & $3.70$ & $5.30$ & $7.06$ & $0.265 \pm 0.041$ & $0.626 \pm 0.367$ & $0.4458 \pm 0.0968$ \\ 
17629 & 2007jw & $30.6364746094$ & $-1.0899255276$ & $0.13690 \pm 0.00007$ & $11.07$ & $11.13$ & $11.13$ & $5.64$ & $7.92$ & $8.67$ & $0.084 \pm 0.028$ & $-0.502 \pm 0.213$ & $-0.1519 \pm 0.0665$ \\ 
17745 & 2007ju & $2.9602687359$ & $-0.3393539488$ & $0.06360 \pm 0.00050$ & $8.87$ & $8.88$ & $8.89$ & $3.26$ & $3.26$ & $3.33$ & $-0.056 \pm 0.028$ & $0.882 \pm 0.274$ & $0.0199 \pm 0.0981$ \\ 
17791 & 2007kp & $332.3733825684$ & $0.7380061746$ & $0.28620 \pm 0.00050$ & $9.30$ & $9.38$ & $9.48$ & $1.99$ & $2.78$ & $3.82$ & $-0.214 \pm 0.068$ & $-0.372 \pm 0.768$ & $0.5136 \pm 0.1888$ \\ 
17801 & 2007ko & $316.0938110352$ & $-0.8984486461$ & $0.20640 \pm 0.00050$ & $11.26$ & $11.34$ & $11.44$ & $4.31$ & $6.29$ & $7.63$ & $0.029 \pm 0.049$ & $-0.306 \pm 0.554$ & $0.2287 \pm 0.1241$ \\ 
17809 & 2007kr & $6.3649182320$ & $-0.8392885327$ & $0.28200 \pm 0.00500$ & $9.66$ & $9.72$ & $9.79$ & $2.02$ & $2.55$ & $3.62$ & $0.033 \pm 0.037$ & $1.674 \pm 0.499$ & $-0.0922 \pm 0.1303$ \\ 
17811 & 2007ix & $12.8806476593$ & $-0.9462078214$ & $0.21320 \pm 0.00050$ & $9.97$ & $10.10$ & $10.21$ & $5.71$ & $8.21$ & $10.21$ & $-0.150 \pm 0.030$ & $0.817 \pm 0.365$ & $-0.0017 \pm 0.0966$ \\ 
17875 & 2007jz & $20.9837265015$ & $1.2550705671$ & $0.23230 \pm 0.00050$ & $10.50$ & $10.62$ & $10.69$ & $4.49$ & $6.49$ & $7.77$ & $-0.086 \pm 0.038$ & $0.753 \pm 0.338$ & $0.1205 \pm 0.0983$ \\ 
17884 & 2007kt & $27.5993213654$ & $1.1723767519$ & $0.23900 \pm 0.00500$ & $10.20$ & $10.24$ & $10.28$ & $4.50$ & $6.02$ & $7.32$ & $-0.147 \pm 0.036$ & $0.179 \pm 0.378$ & $-0.0219 \pm 0.1186$ \\ 
18091 & 2007ku & $23.3678874969$ & $0.5246205926$ & $0.37160 \pm 0.00050$ & $11.00$ & $11.03$ & $11.07$ & $4.99$ & $5.14$ & $5.64$ & $-0.189 \pm 0.070$ & $-0.021 \pm 0.777$ & $0.1106 \pm 0.1888$ \\ 
18241 & 2007ks & $312.3875427246$ & $-0.7619610429$ & $0.09500 \pm 0.01000$ & $9.42$ & $9.48$ & $9.55$ & $3.71$ & $5.52$ & $7.12$ & $-0.080 \pm 0.035$ & $-1.273 \pm 0.195$ & $0.1971 \pm 0.3543$ \\ 
18323 & 2007kx & $3.4286384583$ & $0.6523273587$ & $0.15460 \pm 0.00050$ & $9.32$ & $9.37$ & $9.44$ & $3.61$ & $4.76$ & $6.39$ & $-0.059 \pm 0.029$ & $-0.270 \pm 0.284$ & $0.1280 \pm 0.0778$ \\ 
18375 & 2007lg & $11.5163803101$ & $-0.0106749199$ & $0.11040 \pm 0.00050$ & $10.30$ & $10.43$ & $10.50$ & $4.64$ & $6.60$ & $7.70$ & $0.057 \pm 0.019$ & $0.861 \pm 0.163$ & $-0.1803 \pm 0.0521$ \\ 
18415 & 2007la & $337.4775085449$ & $1.0584667921$ & $0.13070 \pm 0.00050$ & $10.90$ & $10.98$ & $11.05$ & $6.02$ & $8.02$ & $10.15$ & $-0.034 \pm 0.041$ & $-2.093 \pm 0.287$ & $-0.0752 \pm 0.0897$ \\ 
18485 & 2007nu & $47.9590339661$ & $-0.6926384568$ & $0.28200 \pm 0.00050$ & $10.72$ & $10.78$ & $10.83$ & $4.79$ & $6.06$ & $6.73$ & $-0.072 \pm 0.038$ & $1.323 \pm 0.474$ & $-0.0585 \pm 0.1116$ \\ 
18486 & 2007ln & $55.1812210083$ & $1.0045801401$ & $0.24030 \pm 0.00060$ & $9.41$ & $9.50$ & $9.58$ & $2.96$ & $4.61$ & $6.28$ & $-0.109 \pm 0.029$ & $1.008 \pm 0.323$ & $-0.1315 \pm 0.0817$ \\ 
18602 & 2007lo & $338.9836730957$ & $0.6091071367$ & $0.13840 \pm 0.00050$ & $9.20$ & $9.27$ & $9.34$ & $3.39$ & $4.89$ & $6.59$ & $0.068 \pm 0.027$ & $0.812 \pm 0.254$ & $0.0493 \pm 0.0623$ \\ 
18604 & 2007lp & $340.9206848145$ & $0.4205097556$ & $0.17610 \pm 0.00050$ & $11.04$ & $11.10$ & $11.17$ & $7.11$ & $9.11$ & $10.51$ & $0.004 \pm 0.035$ & $-2.395 \pm 0.271$ & $0.1204 \pm 0.0818$ \\ 
18612 & 2007lc & $12.2880029678$ & $0.5966250896$ & $0.11504 \pm 0.00015$ & $11.05$ & $11.05$ & $11.07$ & $5.86$ & $7.15$ & $8.90$ & $0.069 \pm 0.024$ & $-1.218 \pm 0.226$ & $-0.1887 \pm 0.0619$ \\ 
18617 & 2007mw & $345.7612915039$ & $0.8493407369$ & $0.32820 \pm 0.00050$ & $9.83$ & $9.93$ & $10.02$ & $2.50$ & $3.80$ & $5.23$ & $-0.001 \pm 0.060$ & $-0.590 \pm 0.805$ & $-0.1412 \pm 0.1913$ \\ 
18650 & 2007lt & $328.4472045898$ & $0.0150281759$ & $0.11300 \pm 0.00500$ & $8.86$ & $8.91$ & $8.93$ & $2.90$ & $3.05$ & $3.71$ & $-0.076 \pm 0.023$ & $0.821 \pm 0.223$ & $0.2338 \pm 0.1552$ \\ 
18721 & 2007mu & $3.0777626038$ & $-0.0776401758$ & $0.40309 \pm 0.00018$ & $11.16$ & $11.17$ & $11.24$ & $4.73$ & $5.64$ & $6.59$ & $-0.098 \pm 0.074$ & $0.569 \pm 0.931$ & $-0.0820 \pm 0.2083$ \\ 
18749 & 2007mb & $12.5465755463$ & $0.6757113338$ & $0.18940 \pm 0.00050$ & $11.14$ & $11.20$ & $11.23$ & $7.96$ & $9.36$ & $9.96$ & $0.096 \pm 0.040$ & $-1.780 \pm 0.477$ & $-0.1039 \pm 0.1069$ \\ 
18751 & 2007ly & $5.7224216461$ & $0.7759432793$ & $0.07130 \pm 0.00050$ & $10.10$ & $10.12$ & $10.14$ & $9.74$ & $11.34$ & $12.34$ & $0.478 \pm 0.077$ & $-1.967 \pm 0.510$ & $0.2231 \pm 0.2149$ \\ 
18782 & 2007ns & $39.2622032166$ & $-0.8667251468$ & $0.36590 \pm 0.00050$ & $10.94$ & $11.10$ & $11.14$ & $3.81$ & $5.01$ & $5.93$ & $-0.275 \pm 0.075$ & $-0.655 \pm 0.838$ & $0.3879 \pm 0.1999$ \\ 
18890 & 2007mm & $16.4433784485$ & $-0.7594780922$ & $0.06643 \pm 0.00016$ & $10.20$ & $10.28$ & $10.34$ & $6.06$ & $7.80$ & $8.20$ & $0.399 \pm 0.061$ & $-2.995 \pm 0.288$ & $0.1322 \pm 0.1316$ \\ 
18927 & 2007nt & $46.6823501587$ & $-0.7540850639$ & $0.21290 \pm 0.00050$ & $10.40$ & $10.46$ & $10.51$ & $5.42$ & $7.31$ & $8.49$ & $0.159 \pm 0.035$ & $-0.834 \pm 0.389$ & $-0.2880 \pm 0.1001$ \\ 
18940 & 2007sb & $10.3486833572$ & $0.4118011594$ & $0.21230 \pm 0.00050$ & $10.17$ & $10.23$ & $10.28$ & $3.33$ & $4.51$ & $5.97$ & $-0.003 \pm 0.031$ & $-0.878 \pm 0.318$ & $-0.0743 \pm 0.0837$ \\ 
18945 & 2007nd & $10.0779104233$ & $-1.0390836000$ & $0.26330 \pm 0.00050$ & $9.61$ & $9.67$ & $9.75$ & $2.23$ & $3.19$ & $4.46$ & $-0.033 \pm 0.037$ & $-0.088 \pm 0.579$ & $0.1640 \pm 0.1179$ \\ 
18965 & 2007ne & $13.5092248917$ & $1.0689095259$ & $0.20660 \pm 0.00050$ & $10.37$ & $10.47$ & $10.53$ & $3.78$ & $4.31$ & $5.78$ & $-0.124 \pm 0.037$ & $-1.310 \pm 0.353$ & $-0.0385 \pm 0.0875$ \\ 
19002 & 2007nh & $42.6161956787$ & $-0.5511860251$ & $0.26290 \pm 0.00050$ & $10.68$ & $10.84$ & $10.95$ & $2.81$ & $4.96$ & $6.83$ & $-0.097 \pm 0.029$ & $0.263 \pm 0.380$ & $-0.0021 \pm 0.0836$ \\ 
19008 & 2007mz & $331.9632873535$ & $-1.0700660944$ & $0.23220 \pm 0.00050$ & $10.40$ & $10.46$ & $10.50$ & $2.64$ & $3.95$ & $5.60$ & $0.089 \pm 0.038$ & $1.210 \pm 0.478$ & $-0.1803 \pm 0.1001$ \\ 
19027 & 2007my & $328.8840332031$ & $-0.3717949390$ & $0.29320 \pm 0.00050$ & $9.42$ & $9.43$ & $9.46$ & $1.81$ & $1.83$ & $1.87$ & $0.006 \pm 0.044$ & $0.217 \pm 0.608$ & $-0.4353 \pm 0.1387$ \\ 
19029 & 2007lu & $330.3953247070$ & $-0.2568780780$ & $0.31950 \pm 0.00050$ & $9.70$ & $9.73$ & $9.78$ & $1.67$ & $1.70$ & $2.24$ & $-0.127 \pm 0.088$ & $3.565 \pm 0.996$ & $0.7472 \pm 0.2665$ \\ 
19033 & 2007of & $316.2346191406$ & $0.0608703010$ & $0.40470 \pm 0.00050$ & $10.01$ & $10.11$ & $10.21$ & $1.89$ & $3.03$ & $4.48$ & $-0.092 \pm 0.090$ & $0.693 \pm 1.215$ & $0.2684 \pm 0.2699$ \\ 
19067 & 2007oq & $325.6280822754$ & $0.9846492410$ & $0.33910 \pm 0.00050$ & $9.67$ & $9.80$ & $9.91$ & $2.71$ & $4.48$ & $5.82$ & $0.016 \pm 0.059$ & $2.872 \pm 1.342$ & $-0.0072 \pm 0.2755$ \\ 
19149 & 2007ni & $31.4603996277$ & $-0.3325760365$ & $0.19600 \pm 0.00500$ & $9.45$ & $9.49$ & $9.54$ & $2.46$ & $3.00$ & $3.62$ & $0.078 \pm 0.027$ & $1.492 \pm 0.329$ & $-0.3660 \pm 0.1133$ \\ 
19174 & 2007or & $25.6597671509$ & $1.0303381681$ & $0.16640 \pm 0.00050$ & $11.03$ & $11.06$ & $11.10$ & $8.62$ & $10.22$ & $10.72$ & $0.074 \pm 0.032$ & $-1.088 \pm 0.305$ & $-0.2163 \pm 0.0798$ \\ 
19211 & 2007oh & $313.1539306641$ & $-0.4541013539$ & $0.41990 \pm 0.00050$ & $10.29$ & $10.40$ & $10.52$ & $2.10$ & $3.73$ & $5.51$ & $-0.107 \pm 0.087$ & $0.702 \pm 1.220$ & $0.2239 \pm 0.2425$ \\ 
19230 & 2007mo & $332.8909912109$ & $0.7647492290$ & $0.22150 \pm 0.00050$ & $10.32$ & $10.42$ & $10.47$ & $4.03$ & $6.03$ & $8.63$ & $0.145 \pm 0.050$ & $-1.493 \pm 0.581$ & $-0.3277 \pm 0.1387$ \\ 
19282 & 2007mk & $359.0729980469$ & $-0.5038936734$ & $0.18641 \pm 0.00016$ & $8.54$ & $8.59$ & $8.64$ & $2.42$ & $2.46$ & $3.00$ & $-0.107 \pm 0.023$ & $0.530 \pm 0.279$ & $-0.0657 \pm 0.0593$ \\ 
19341 & 2007nf & $15.8608398438$ & $0.3316199183$ & $0.22800 \pm 0.00500$ & $10.99$ & $11.03$ & $11.07$ & $7.96$ & $9.96$ & $10.58$ & $0.071 \pm 0.052$ & $-1.940 \pm 0.488$ & $-0.1392 \pm 0.1435$ \\ 
19353 & 2007nj & $43.1132774353$ & $0.2517381907$ & $0.15395 \pm 0.00011$ & $10.83$ & $10.86$ & $10.86$ & $6.77$ & $8.50$ & $8.50$ & $0.059 \pm 0.023$ & $0.863 \pm 0.308$ & $-0.0349 \pm 0.0665$ \\ 
19425 & 2007ow & $323.5083923340$ & $-0.7406359315$ & $0.21160 \pm 0.00050$ & $10.53$ & $10.60$ & $10.68$ & $5.13$ & $8.23$ & $10.13$ & $0.210 \pm 0.060$ & $-1.419 \pm 0.558$ & $-0.0122 \pm 0.1373$ \\ 
19543 & 2007oj & $357.9083862305$ & $0.2798276842$ & $0.12300 \pm 0.00500$ & $8.78$ & $8.84$ & $8.90$ & $3.16$ & $4.03$ & $5.30$ & $-0.007 \pm 0.026$ & $-1.104 \pm 0.235$ & $0.1811 \pm 0.1463$ \\ 
19596 & 2007po & $53.8846893311$ & $0.7037985921$ & $0.29200 \pm 0.00500$ & $9.69$ & $9.78$ & $9.88$ & $2.38$ & $3.50$ & $4.94$ & $-0.033 \pm 0.042$ & $0.986 \pm 0.638$ & $0.0210 \pm 0.1451$ \\ 
19604 & 2007oi & $5.3261027336$ & $1.0737973452$ & $0.29600 \pm 0.00500$ & $10.07$ & $10.17$ & $10.25$ & $2.51$ & $4.26$ & $6.08$ & $0.172 \pm 0.068$ & $1.950 \pm 1.012$ & $0.3947 \pm 0.2209$ \\ 
19626 & 2007ou & $35.9277305603$ & $-0.8264662623$ & $0.11321 \pm 0.00005$ & $10.27$ & $10.30$ & $10.36$ & $3.71$ & $4.73$ & $6.24$ & $0.329 \pm 0.034$ & $1.744 \pm 0.461$ & $0.4453 \pm 0.0815$ \\ 
19632 & 2007ov & $40.2866287231$ & $0.1442469060$ & $0.31530 \pm 0.00050$ & $10.97$ & $11.06$ & $11.15$ & $4.25$ & $7.12$ & $8.72$ & $-0.025 \pm 0.046$ & $-0.052 \pm 0.596$ & $-0.0682 \pm 0.1344$ \\ 
19658 & 2007ot & $8.9032306671$ & $-0.2325988412$ & $0.20000 \pm 0.00050$ & $8.72$ & $8.78$ & $8.87$ & $2.37$ & $2.97$ & $4.12$ & $-0.093 \pm 0.030$ & $-0.401 \pm 0.350$ & $0.0771 \pm 0.0756$ \\ 
19757 & 2007oy & $349.4814147949$ & $1.2236189842$ & $0.40300 \pm 0.00500$ & $10.26$ & $10.39$ & $10.56$ & $1.98$ & $3.51$ & $5.47$ & $-0.087 \pm 0.087$ & $0.144 \pm 1.079$ & $-0.3474 \pm 0.2605$ \\ 
19775 & 2007pc & $318.9561462402$ & $0.6512132883$ & $0.13790 \pm 0.00050$ & $10.78$ & $10.88$ & $10.97$ & $4.78$ & $6.75$ & $8.32$ & $0.119 \pm 0.027$ & $-0.621 \pm 0.314$ & $-0.0728 \pm 0.0694$ \\ 
19794 & 2007oz & $359.3190917969$ & $0.2484871745$ & $0.29730 \pm 0.00018$ & $11.16$ & $11.41$ & $11.44$ & $4.28$ & $9.38$ & $9.88$ & $0.068 \pm 0.094$ & $-2.117 \pm 0.938$ & $0.0166 \pm 0.2460$ \\ 
19818 & 2007pe & $35.2665252686$ & $0.4965370297$ & $0.30440 \pm 0.00050$ & $10.14$ & $10.22$ & $10.30$ & $2.54$ & $4.19$ & $5.54$ & $-0.056 \pm 0.040$ & $0.613 \pm 0.556$ & $-0.2044 \pm 0.1116$ \\ 
19913 & 2007qf & $333.7622070312$ & $-0.3417298794$ & $0.20380 \pm 0.00050$ & $9.79$ & $9.83$ & $9.85$ & $2.41$ & $2.89$ & $3.57$ & $-0.056 \pm 0.028$ & $0.245 \pm 0.457$ & $0.1322 \pm 0.0856$ \\ 
19940 & 2007pa & $315.3935546875$ & $-0.2687674761$ & $0.15710 \pm 0.00080$ & $8.74$ & $8.85$ & $8.98$ & $3.26$ & $4.60$ & $6.59$ & $0.019 \pm 0.025$ & $1.099 \pm 0.340$ & $-0.3307 \pm 0.0676$ \\ 
19969 & 2007pt & $31.9098148346$ & $-0.3240273297$ & $0.17529 \pm 0.00010$ & $10.31$ & $10.32$ & $10.34$ & $2.49$ & $2.53$ & $3.07$ & $0.034 \pm 0.025$ & $-0.485 \pm 0.345$ & $-0.1577 \pm 0.0711$ \\ 
19990 & 2007ps & $34.8060150146$ & $-0.3845337927$ & $0.24600 \pm 0.00500$ & $10.52$ & $10.59$ & $10.64$ & $6.38$ & $8.78$ & $10.38$ & $-0.038 \pm 0.044$ & $-1.164 \pm 0.491$ & $0.0404 \pm 0.1339$ \\ 
20040 & 2007rf & $328.8794555664$ & $0.8150795698$ & $0.28800 \pm 0.00050$ & $10.27$ & $10.38$ & $10.47$ & $3.97$ & $6.37$ & $8.53$ & $-0.097 \pm 0.046$ & $0.829 \pm 0.731$ & $-0.1062 \pm 0.1240$ \\ 
20048 & 2007pq & $339.3081054688$ & $0.7363132834$ & $0.18550 \pm 0.00050$ & $10.71$ & $10.80$ & $10.87$ & $7.01$ & $8.01$ & $11.01$ & $0.050 \pm 0.040$ & $-0.939 \pm 0.567$ & $0.0230 \pm 0.1211$ \\ 
20064 & 2007om & $358.5862731934$ & $-0.9172353745$ & $0.10503 \pm 0.00018$ & $11.16$ & $11.16$ & $11.27$ & $4.69$ & $4.69$ & $6.85$ & $0.107 \pm 0.023$ & $0.408 \pm 0.322$ & $-0.2490 \pm 0.0619$ \\ 
20106 & 2007pr & $346.5540771484$ & $0.3289288580$ & $0.33300 \pm 0.00500$ & $10.12$ & $10.20$ & $10.26$ & $3.05$ & $5.27$ & $6.52$ & $-0.042 \pm 0.084$ & $-0.503 \pm 0.905$ & $0.0408 \pm 0.2340$ \\ 
20111 & 2007pw & $354.3940734863$ & $0.2474300116$ & $0.24500 \pm 0.00500$ & $10.78$ & $10.87$ & $10.93$ & $5.90$ & $8.39$ & $8.94$ & $0.015 \pm 0.048$ & $-0.362 \pm 0.810$ & $-0.0164 \pm 0.1594$ \\ 
20184 & 2007qn & $359.7885131836$ & $1.1585552692$ & $0.32400 \pm 0.00050$ & $9.53$ & $9.65$ & $9.75$ & $2.21$ & $3.38$ & $4.62$ & $0.029 \pm 0.082$ & $-1.095 \pm 1.398$ & $-0.2104 \pm 0.2047$ \\ 
20227 & 2007qi & $349.1200561523$ & $-0.0988994613$ & $0.27640 \pm 0.00050$ & $10.64$ & $10.71$ & $10.76$ & $5.08$ & $7.48$ & $8.64$ & $-0.134 \pm 0.057$ & $-1.647 \pm 0.713$ & $0.0382 \pm 0.1498$ \\ 
20350 & 2007ph & $312.8067932129$ & $-0.9577776194$ & $0.12946 \pm 0.00018$ & $11.19$ & $11.29$ & $11.34$ & $5.03$ & $8.14$ & $9.63$ & $0.213 \pm 0.054$ & $-3.029 \pm 0.845$ & $-0.0720 \pm 0.1358$ \\ 
20364 & 2007qo & $25.7565631866$ & $-0.9451811910$ & $0.21810 \pm 0.00090$ & $10.31$ & $10.37$ & $10.41$ & $2.76$ & $3.60$ & $5.70$ & $0.062 \pm 0.039$ & $0.830 \pm 0.964$ & $0.0192 \pm 0.1404$ \\ 
20376 & 2007re & $319.3955078125$ & $-0.5239647627$ & $0.21090 \pm 0.00050$ & $10.61$ & $10.69$ & $10.74$ & $5.74$ & $9.14$ & $10.24$ & $0.164 \pm 0.045$ & $-1.096 \pm 0.874$ & $-0.3531 \pm 0.1499$ \\ 
20821 & 2007rk & $55.5723648071$ & $1.0622460842$ & $0.19590 \pm 0.00050$ & $10.58$ & $10.63$ & $10.68$ & $2.94$ & $4.02$ & $5.60$ & $0.285 \pm 0.057$ & $-1.504 \pm 1.272$ & $0.0002 \pm 0.2060$ \\ 

\enddata
\tablenotetext{a}{Redshift error $\geq 0.005$ corresponds to a redshift from the SN spectrum; 
redshift error $\leq 0.0005$ corresponds to a redshift from the host galaxy.}

\end{deluxetable}
\clearpage
\end{landscape}

\end{document}